\documentclass[fleqn,usenatbib]{mnras}

\usepackage{newtxtext,newtxmath}

\usepackage[T1]{fontenc}

\DeclareRobustCommand{\VAN}[3]{#2}
\let\VANthebibliography\thebibliography
\def\thebibliography{\DeclareRobustCommand{\VAN}[3]{##3}\VANthebibliography}

\usepackage[normalem]{ulem}  
\newcommand\redout{\bgroup\markoverwith
{\textcolor{red}{\rule[0.5ex]{2pt}{0.8pt}}}\ULon}

\usepackage{graphicx}	
\usepackage{amsmath}	

\usepackage{longtable}
\usepackage{lscape}
\usepackage{multirow}

\newcommand{\ha}{H$\alpha$} 
\newcommand{\hb}{H$\beta$}

\newcommand{\sulfurii}{[S\,{\sc~ii}]}

\newcommand{\nitrogen}{[N\,{\sc~ii}]}

\newcommand{\oxygeniii}{[O\,{\sc~iii}]}

\newcommand{\oxygeni}{[O\,{\sc~i}]}
\newcommand{\oxygenii}{[O\,{\sc~ii}]}

\newcommand{\sulfurt}{[S\,{\sc~ii}]}

\usepackage{tikz,xcolor,hyperref}

\definecolor{lime}{HTML}{A6CE39}
\DeclareRobustCommand{\orcidicon}{
	\begin{tikzpicture}
	\draw[lime, fill=lime] (0,0) 
	circle [radius=0.16] 
	node[white] {{\fontfamily{qag}\selectfont \tiny ID}};
	\draw[white, fill=white] (-0.0625,0.095) 
	circle [radius=0.007];
	\end{tikzpicture}
	\hspace{-2mm}
}

\foreach \x in {A, ..., Z}{\expandafter\xdef\csname orcid\x\endcsname{\noexpand\href{https://orcid.org/\csname orcidauthor\x\endcsname}
			{\noexpand\orcidicon}}
}

\title[Low-ionization structures in planetary nebulae]{Low-ionization structures in planetary nebulae -- III. The statistical analysis of physico-chemical parameters and excitation mechanisms}

\author[Mari, Akras and Gon\c{c}alves]{
M. Bel\'en Mari$^{1\orcidB{}}$ \thanks{E-mail: mbmari@astro.ufrj.br},
Stavros Akras$^{2\orcidA{}}$ and Denise R. Gon\c{c}alves$^{1}$
\\
$^{1}$Observat\'orio do Valongo, Universidade Federal do Rio de Janeiro, Ladeira Pedro Antonio 43, Rio de Janeiro 20080-090, Brazil\\
$^{2}$Institute for Astronomy, Astrophysics, Space Applications and Remote Sensing, National Observatory of Athens, Penteli GR 15236, Greece
}

\date{Accepted XXX. Received YYY; in original form ZZZ}

\pubyear{2021}

\begin{document}
\label{firstpage}
\pagerange{\pageref{firstpage}--\pageref{lastpage}}
\maketitle
\newcommand{\stavros}[1]{{\color{teal}  #1}}
\newcommand{\denise}[1]{{\color{red}  #1}}
\newcommand{\belu}[1]{{\color{violet}  #1}}

\begin{abstract}
Nearly 30 years after the first detailed studies of low-ionization structures (LISs) in planetary nebulae (PNe), we perform a statistical analysis of their physical, chemical and excitation properties, by collecting published data in the literature. The analysis was made through the contrast between LISs and high-ionization structures – rims or shells – for a large sample of PNe, in order to highlight significant differences between these structures. Our motivation was to find robust results based on the largest sample of LISs gathered so far. (i) Indeed, LISs have lower electron densities (N$_{e}$[S{\sc~ii}]) than the rims/shells. (ii) The nitrogen electron temperatures (T$_{e}$[N{\sc~ii}]) are similar between the two groups, while a bimodal distribution is observed for the T$_{e}$ based on [O{\sc~iii}] of the rims/shells, although the high- and low-ionization structures have T$_{e}$[O~{\sc iii}] of similar median values. (iii) No significant variations are observed in total abundances of He, N, O, Ne, Ar, Cl and S between the two groups. (iv) Through the analysis of several diagnostic diagrams, LISs are separated from rims/shells in terms of excitation. From two large grids of photoionization and shock models, we show that there is an important overlap between both mechanisms, particularly when low-ionization line-ratios are concerned. We found a good tracer of high-velocity shocks, as well as an indicator of high- and low-velocity shocks that depends on temperature-sensitive line ratios. In conclusion, both excitation mechanisms could be present, however shocks cannot be the main source of excitation for most of the LISs of PNe.

\end{abstract}




\section{Introduction}

This is the third of a series of papers carrying out an optical spectroscopic study of low-ionization structures \citep[LISs; ][]{2001ApJ...547..302G} and their host planetary nebulae (PNe). In Paper~I \citep{Akras2016} and II \citep*[][]{2023MNRAS.518.3908M}, the spectroscopic study of 5 and 6 PNe, respectively, were presented. These papers completed the analysis of the sample whose initial data were published in \citet[][]{Goncalves2003,Goncalves2004,2009MNRAS.398.2166G}. The spectroscopic analysis was carried out for two different groups of nebular components: the high-ionization ones -- rims or shells -- and low-ionization structures. 

Overall, PNe have large-scale structures such as rims and shells, 
bright in the light of hydrogen and helium recombination lines, as well as in the forbidden [O~{\sc~iii}] lines. The formation of the rims and shells in PNe is relatively well understood \citep[see][for a review]{2002ARA&A..40..439B}. On somewhat smaller scales, they can present LISs, 
visible primarily in low-ionization species such as [N~{\sc ii}], [S~{\sc ii}], [O~{\sc ii}] and also [O~{\sc i}] \citep[e.g.][]{1993ApJ...411..778B,1996A&A...313..913C,2001ApJ...547..302G}. The origin of these 
small-scale structures still remain an open question in the field of photoionized nebula. Several studies have been performed since their earlier report \citep[][]{1987AJ.....94..671B}, using either imaging \citep{1987AJ.....94..671B,1992A&AS...96...23S,1996iacm.book.....M,1996A&A...313..913C,1998AJ....116..360B,1999A&AS..136..145G} or spectroscopic data \citep[][]{1994ApJ...424..800B,1997ApJ...487..304H,Goncalves2003,Goncalves2004,2009MNRAS.398.2166G,Akras2016,2016AJ....151...38D,Ali2017,2020A.A...634A..47M,2021arXiv210505186M,Akras2022,2023MNRAS.518.3908M}. These spectroscopic studies about LISs and their host PNe came, independently, to the conclusion that LISs are characterized by lower -- or at most equal -- electron density (N$_e$) than the surrounding gas (rims and shells), while the electron temperature (T$_e$) and the chemical composition of rims, shells and LISs appear to be the same.

Such low N$_e$ in LISs contradicts the formation mechanisms proposed to explain these micro-structures, as most of the theoretical reasoning and models consider them as dense structures moving in a tenuous environment \citep[][]{2001ApJ...556..823S,2008A&A...489.1141R,2020ApJ...889...13B}. The 
fact that molecular hydrogen (H$_2$) emission from the cometary knots in the Helix nebula \citep[][]{2009ApJ...700.1067M} was known for a while 
led Gon\c calves and collaborators to proposed that LISs, other than the cometary knots, may also contain molecular gas and dust \citep[][]{2009MNRAS.398.2166G}. Recent studies focused on the near-infrared ro-vibrational H$_2$ line centred at 2.12~$\mu$m have unveiled the H$_2$ counterpart of several LISs \citep[][]{2015MNRAS.452.2445F,Akras2017,2018ApJ...859...92F,Akras2020b}. These H$_2$ condensations have a size around 2-5$\times$10$^{16}$~cm, while the host PNe are relatively young, $<$2000~years \citep{Akras2020c}. These findings imply the presence of high-density gas, enough to shield the molecular matter from the central star UV radiation and prevent its dissociation, as predicted by LISs' formation models \citep[e.g.][]{2020ApJ...889...13B}.

The dominant excitation mechanism in LISs can be either photoionization from the UV radiation of the central stars or shock interaction with the other nebular components or the circumstellar medium. Both mechanisms are supported by the 
enhanced low-ionization line ratios (e.g. \nitrogen~/\ha, \sulfurt/\ha, \oxygeni/\ha, etc.) observed in LISs \citep[using diagnostic techniques as, e.g.][]{Sabbadin1977,2008A&A...489.1141R}.  

In the context of an overall view of LISs in PNe, publicly available spectroscopic 
results, from the literature, for PNe with LISs, were gathered with the intention of carrying out the first statistical analysis of their physical, chemical and excitation mechanisms, to identify potential patterns and trends.

The paper is organized as follows: the data sample gathered from the literature and their visualization are presented in section~\ref{sec2}. The results of our statistical analysis are presented in Section~\ref{sec3}. In Section~\ref{sec4}, we discuss the predictions from photoionization and shocks models. Diagnostic diagrams for the separation of photoionized and shocks-heated gases are discussed. Overall 
discussions and conclusions appear in Section~\ref{sec5} and \ref{sec6}, respectively.

\section{Data sample and visualization}
\label{sec2}
A statistical study of the physical, chemical and excitation mechanisms of PNe and their LISs is 
missing. To solve this problem, in this work, we have gathered  spectroscopic information for LISs and their host PNe available in the literature.
Table~\ref{tab:datasample} lists 
these objects and the references from which we collected the data. 
In total, our sample consists of 33 
PNe, with 88 Rims/Shells 
and 104 LISs, the largest and most complete sample 
analysed this far. LISs refers to -- generally small -- structures, bright in low-ionization lines, with the appearance of knots or filaments. Rims and shells, on the other hand, are of larger scales, much higher in ionization and result from the interacting AGB/post-AGB stellar winds,  photoionized by the central star radiation \citep[e.g.][]{1987AJ.....94..671B}. 

It is important to point out the fact that, since the data collected were published over $\sim$30 years by different authors, it is not homogeneous in terms of atomic data, excitation curves, ionization correction factors for total abundances, etc. The line ratios (already corrected by extinction), as well as the physico-chemical properties, were taken without further manipulation. The exception are  \citet{Goncalves2003,Goncalves2004,2009MNRAS.398.2166G} objects, for which we applied the c$_{\beta}$ correction using their reported values of extinction. 
\begin{table}
\caption{PNe with LISs, from the literature.}
\label{tab:datasample}
\begin{tabular}{lccl}
\hline

Name &  Rims/Shells & LISs & References \\
\hline            
NGC~6543$^{\dag}$ & 2 &   1   & \cite{1994ApJ...424..800B} \\
NGC~6826          & 1 &   1   & \cite{1994ApJ...424..800B} \\
NGC~7009$^{\dag}$ & 2 &   1   & \cite{1994ApJ...424..800B} \\
Hb~4              & 2 &   2   & \cite{1997ApJ...487..304H} \\
IC~4634$^{\dag}$  & 3 &   2   & \cite{1997ApJ...487..304H} \\
NGC~6369          & 2 &   2   & \cite{1997ApJ...487..304H} \\
NGC~7354$^{\dag}$ & 4 &   2   & \cite{1997ApJ...487..304H} \\
M~2-48            & 1 &   2   & \citet{2002AA...388..652L} \\
NGC~7009$^{\dag}$ & 2 &   6   & \cite{Goncalves2003}       \\
K~4-47            & 1 &   2   & \cite{Goncalves2004}       \\
NGC~7662$^{\dag}$ & 1 &   6   & \citet{2004A.A...422..963P} \\
IC~4634$^{\dag}$  & 3 &   2   & \citet{2008ApJ...683..272G} \\
He~1-1            & 1 &   2   & \cite{2009MNRAS.398.2166G} \\
IC~2149           & 1 &   2   & \cite{2009MNRAS.398.2166G} \\
KjPn~8            & - &   2   & \cite{2009MNRAS.398.2166G} \\
NGC~7662$^{\dag}$ & 7 &   2   & \cite{2009MNRAS.398.2166G} \\
NGC~7354$^{\dag}$ & 11 &  10  & \citet{2010AJ....139.1426C} \\
The Necklace      & 1 &   1   & \citet{2011MNRAS.410.1349C} \\
ETHOS~1           & 2 &   3   & \citet{2011MNRAS.413.1264M} \\
NGC~3242          & 2 &   5   & \citet{2013A.A...560A.102M} \\
Hu~1-2            & 2 &   1   & \citet{2015MNRAS.452.2445F} \\
IC~4846           & 1 &   2   & \citet{Akras2016}  \\
Wray~17-1         & 6 &   6   & \citet{Akras2016}  \\
K~1-2             & - &   5   & \citet{Akras2016}  \\
NGC~6891          & 6 &   2   & \citet{Akras2016}  \\
NGC~6572          & 2 &   4   & \citet{Akras2016}  \\
M~2-42            & 1 &   2   & \cite{2016AJ....151...38D} \\
NGC~5307          & - &   3   & \cite{Ali2017} \\
IC~2553           & - &   2   & \cite{Ali2017} \\
PB~6              & - &   1   & \cite{Ali2017} \\
NGC~3132          & - &   2   & \citet{2020A.A...634A..47M} \\
IRAS~18061–2505   & 2 &   2   & \cite{2021arXiv210505186M}                \\
IC~4593           & 4 &   3   & \citet{2023MNRAS.518.3908M} \\
Hen~2-186         & 1 &   2   & \citet{2023MNRAS.518.3908M} \\
Hen~2-429         & 1 &   2   & \citet{2023MNRAS.518.3908M} \\
NGC~3918          & 3 &   5   & \citet{2023MNRAS.518.3908M} \\
NGC~6543$^{\dag}$ & 8 &   2   & \citet{2023MNRAS.518.3908M} \\
NGC~6905          & 2 &   2   & \citet{2023MNRAS.518.3908M} \\
\hline
Total             & 88 & 104  &          \\
\hline
\end{tabular}

Note: $^{\dag}$ These PNe, NGC~6543, NGC~7009, IC~4634, NGC~7354 and NGC~7662, are repeated because the structures studied in each work are, in general, different.\\
\end{table}

An integrated way to visualize and explore this sample is through 
the use of \textit{violinplots} in conjunction with the statistical analysis of the \textit{boxplots}, embedded in the former plots. This type of representation allows for different shapes of the distributions and makes clear the presence of clustering, such as bimodalities (see Appendix~\ref{apA}), as they use a kernel density distribution. To detect significant differences between the samples, the use of \textit{notches} is of great help, as they represent the 95\% confidence interval (CI) for the median \citep[]{10.2307/2685478}. When the notches of the distributions of two groups do not overlap, we can safely conclude that the samples indeed differ \citep{2017..book.....C}. Considering that the aim of this work is to find different trends between Rims/Shells and LISs in PNe, we work with violinplots throughout this study.

\section{Results}
\label{sec3}

Here, we perform a statistical analysis of the N$_e$ (Section~\ref{sec_ne}) and T$_e$ (Section~\ref{sec_te}) obtained from the \sulfurt, \nitrogen~ and \oxygeniii~ diagnostic lines. Then a similar 
analysis is carried out for the chemical abundances of the host PNe and LISs (Section~\ref{sec_abun}). We finish the presentation of the results exploring a few characteristic emission-line ratios, which concern the probable excitation mechanism present in the different structures of the PNe studied (Section~\ref{sec_exc}).

\begin{figure*}
    \centering
    \includegraphics[width=0.8\textwidth]{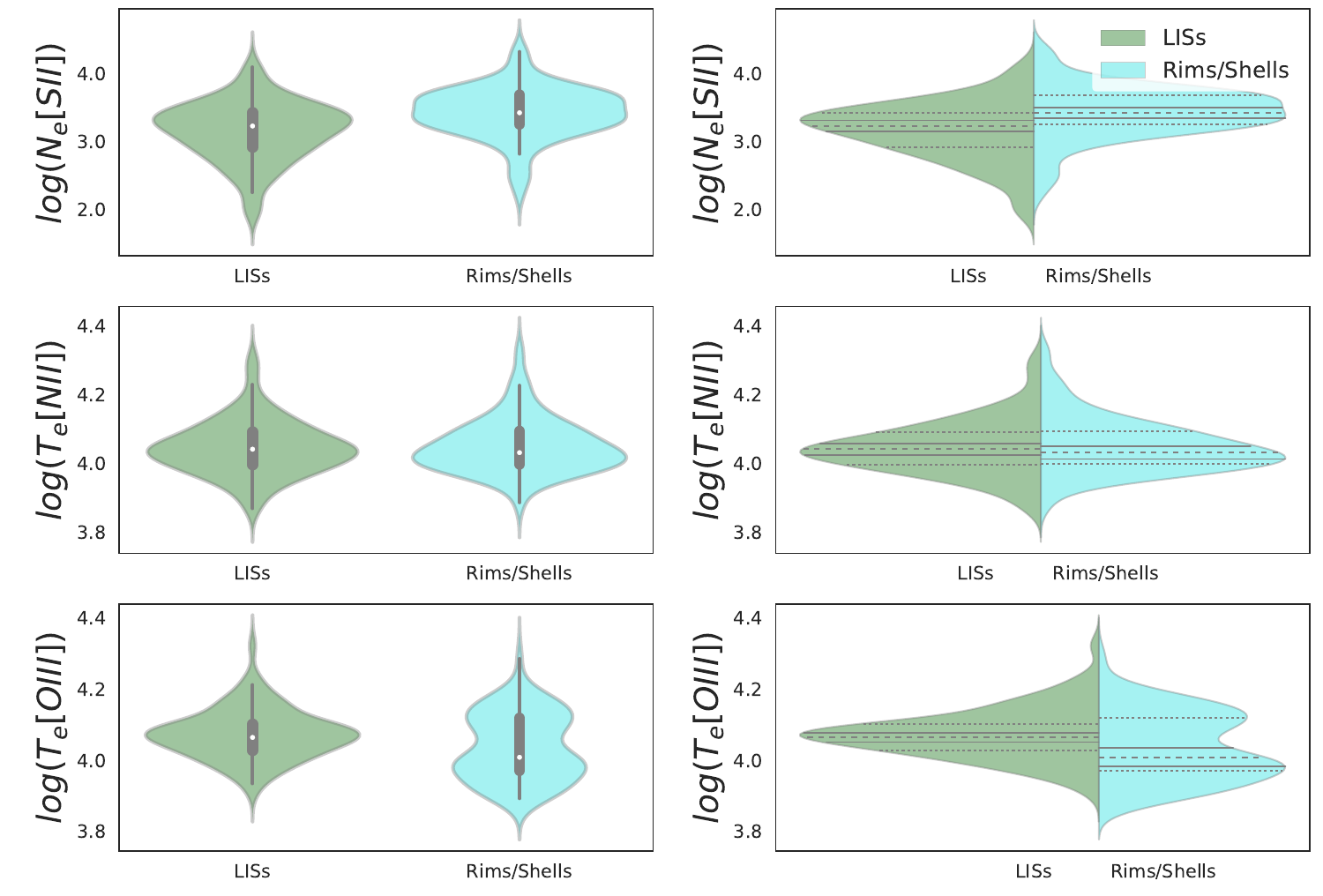}
    \caption{Violinplots showing the electron temperature and density for the two groups: Rims/Shells (in cyan) and LISs (in green). \textit{Left panels}: comparison between the two components with violinplots with boxplots inside. \textit{Right panels}: comparison between LISs and Rims/Shells using split violinplots with the median (Q2), 25th and 75th quartiles (Q1 and Q3) shown by dashed and dotted lines, respectively, and the notches represented by solid lines.} 
    \label{fig:violinplotTN}
\end{figure*}
\begin{table*}
\caption{Group properties for the samples of Rims/Shells and LISs. The notches correspond to the approximated 95 per cent CIs. The last two rows show the number of outliers and sample size, respectively.}
\label{tab:violinplot}
\begin{tabular}{l|cccccc}
\hline
            & \multicolumn{2}{c}{log(N$_{e}$[S~{\sc~ii}])} & \multicolumn{2}{|c}{log(T$_{e}$[N~{\sc~ii}])}     & \multicolumn{2}{c}{log(T$_{e}$[O~{\sc~iii}])} \\

\hline            
            & Rims/Shells       & LISs        & Rims/Shells       & LISs        & Rims/Shells       & LISs     \\
Mean        & 3.4308            & 3.1596      & 4.0498            & 4.0447      & 4.0344            & 4.0726   \\
Median      & 3.4249            & 3.2326      & 4.0314            & 4.0414      & 4.0086            & 4.0645   \\
Lower notch & 3.3477            & 3.1510      & 4.0125            & 4.0246      & 3.9826            & 4.0516   \\
Upper notch & 3.5021            & 3.3143      & 4.0503            & 4.0582      & 4.0346            & 4.0773   \\
IQR         & 0.4370            & 0.5149      & 0.0962            & 0.0964      & 0.1470            & 0.0747   \\
Q1          & 3.2576            & 2.9177      & 3.9976            & 3.9956      & 3.9696            & 4.0273   \\
Q3          & 3.6946            & 3.4326      & 4.0938            & 4.0920      & 4.1166            & 4.1021   \\
\#Outliers  & 4                 & 3           & 2                 & 2           & 0                 & 3        \\
\#Sample    & 79                & 98          & 64                & 81          & 79                & 83       \\
\hline

\end{tabular} 
\end{table*}

\subsection{Electron density} \label{sec_ne} 

In the top panels of Figure~\ref{fig:violinplotTN}, we present the N$_e$ for the two subsets of Rims/Shells (cyan) and LISs (green). The size of the thick vertical black lines at the centre of the violinplots represent the  interquartile range (IQR), whereas the white dot corresponds to the median value of each data set. The distributions are found to be similar in both groups, with comparable widths. As expected from previous studies, the group of LISs (sample size equal to 98) clearly shows a peak at lower densities compared to the group of Rims/Shells (80), whereas there is a small number of LISs, Rims and Shells, which exhibit N$_e$ close to or even higher than 10$^4$~cm$^{-3}$. Likewise, there is a small number of Rims/Shells with N$_e$ lower than 10$^{2.5}$~cm$^{-3}$. The wing of the distribution may indicate measurements with high uncertainties and outliers (see Table~\ref{tab:violinplot}).

In Table~\ref{tab:violinplot}, we list the statistical quantifies for both groups. The median values of log(N$_e$) are 3.42 ($\sim$2700~cm$^{-3}$) and 3.23 ($\sim$1700~cm$^{-3}$) for the Rims/Shells and LISs, respectively. Taking into account the lower and upper notches, it is clear that there is no overlap between the two groups, with a $\sim$95\% of CI. Therefore, LISs represent a statistically different sample than the Rims/Shells in terms of electron density.

\subsection{Electron temperature} \label{sec_te}

The middle panels of Fig.~\ref{fig:violinplotTN} show the T$_e$ from the \nitrogen~emission-lines. The T$_e$\nitrogen~distribution is nearly similar for both groups. 
The median value of log(T$_e$\nitrogen) is almost the same for LISs and Rims/Shells, being 4.04 ($\sim$11000~K) and 4.03 ($\sim$10700~K), respectively. The upper and lower notches of the two type of structures (see Table~\ref{tab:violinplot}) allow the clear conclusion that both groups are identical in terms of T$_e$\nitrogen. On the other hand, looking at the violinplots of T$_e$\oxygeniii~(bottom-left panel in Fig.~\ref{fig:violinplotTN}) we note that the group of Rims/Shells displays a bimodal distribution with peaks at $\sim$3.97 and $\sim$4.12, while LISs show a nearly bell-shell distribution with a peak at $\sim$4.07. Scrutinizing the results of log(T$_e$\oxygeniii), we also notice that the median value of Rims/Shells is lower than the values obtained from LISs (see the white dots Fig.~\ref{fig:violinplotTN}, in bottom-left panel). In particular, LISs have a median log(T$_e$\oxygeniii) value of 4.06 ($\sim$11500~K) and Rims/Shells of 4.01 ($\sim$10200~K). This difference is significant, and we could argue that the two groups are statistically different in terms of T$_e$\oxygeniii~as there is no overlap of their notches (see Table~\ref{tab:violinplot}). However, we should also note that the peak of log(T$_e$\oxygeniii) for the LISs is similar to the valley of the bimodality present in the distribution of the Rims/Shells -- whose peaks approximately coincide with the Q1 and Q3 of its distribution. This result shows that a significant number of LISs has higher T$_e$\oxygeniii~than Rims/Shells, which could be interpreted as an extra excitation mechanism in LISs.

The right panel of Fig.~\ref{fig:violinplotTN} illustrate the \textit{split} distributions (i.e. one-half of the violinplots) of N$_e$ and T$_e$ side-by-side, for a direct comparison. 
The median, notches and IQR parameters for the two groups, as well as the size of the samples, are listed in Table~\ref{tab:violinplot}. The straightforward results from the table and plots is that 
LISs and Rims/Shells have different properties in terms of N$_e$\sulfurt~and T$_e$\oxygeniii~ 
and behave similarly when T$_e$\nitrogen is concerned.

\subsection{Abundances} \label{sec_abun} 

\begin{figure*}
    \centering
    \includegraphics[width=0.498\textwidth]{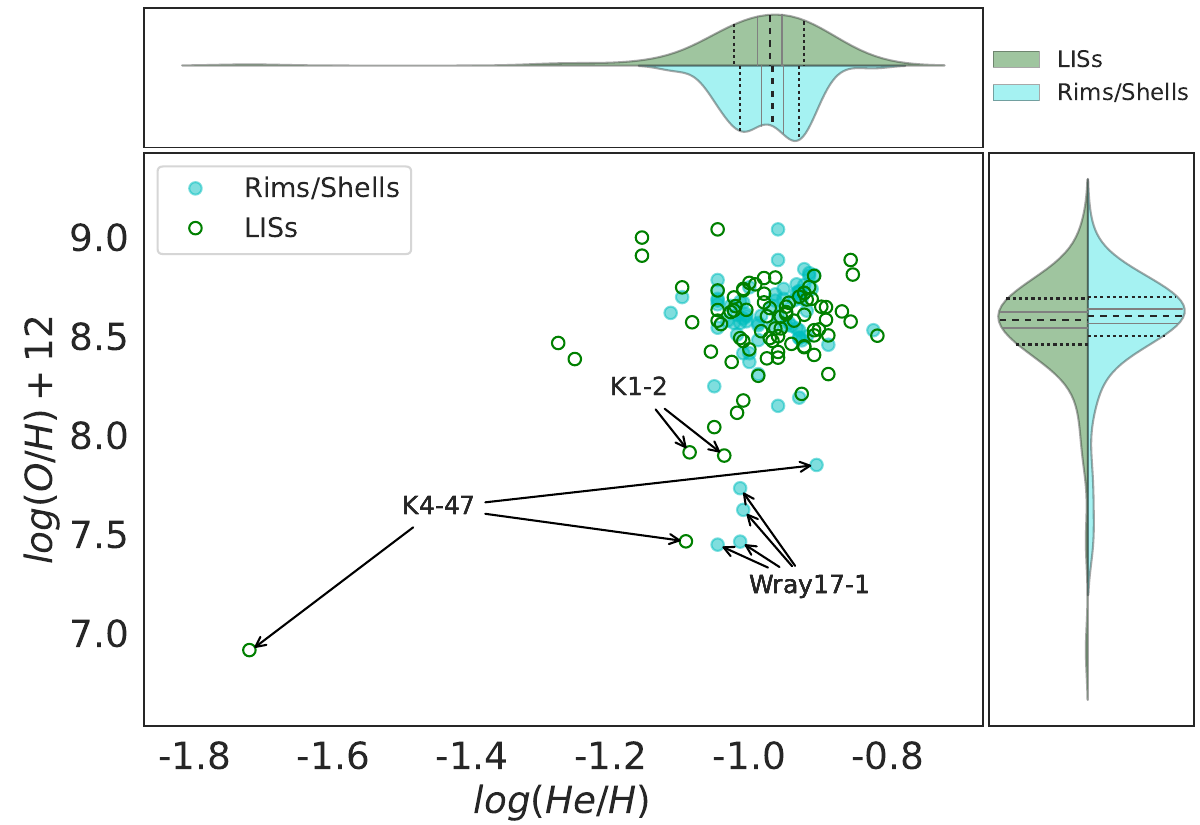}    
    \includegraphics[width=0.498\textwidth]{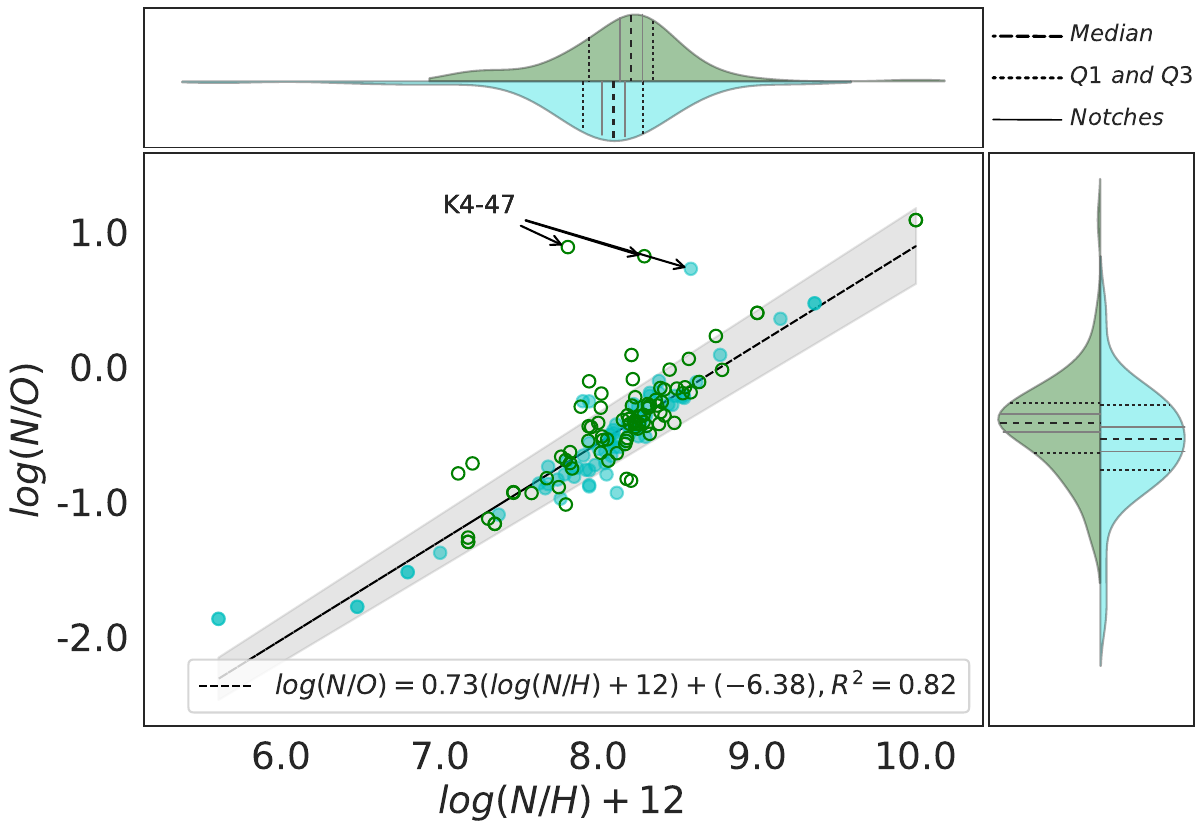}
    \includegraphics[width=0.498\textwidth]{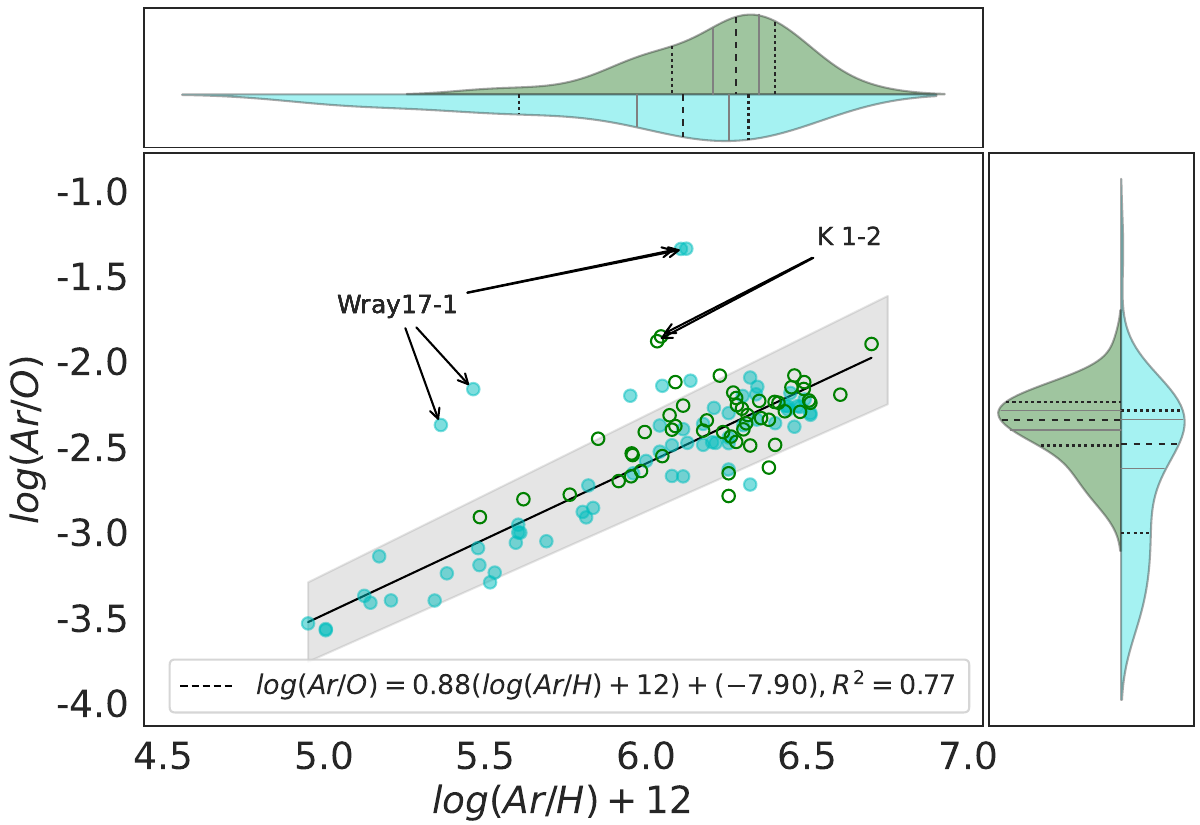}
    \includegraphics[width=0.498\textwidth]{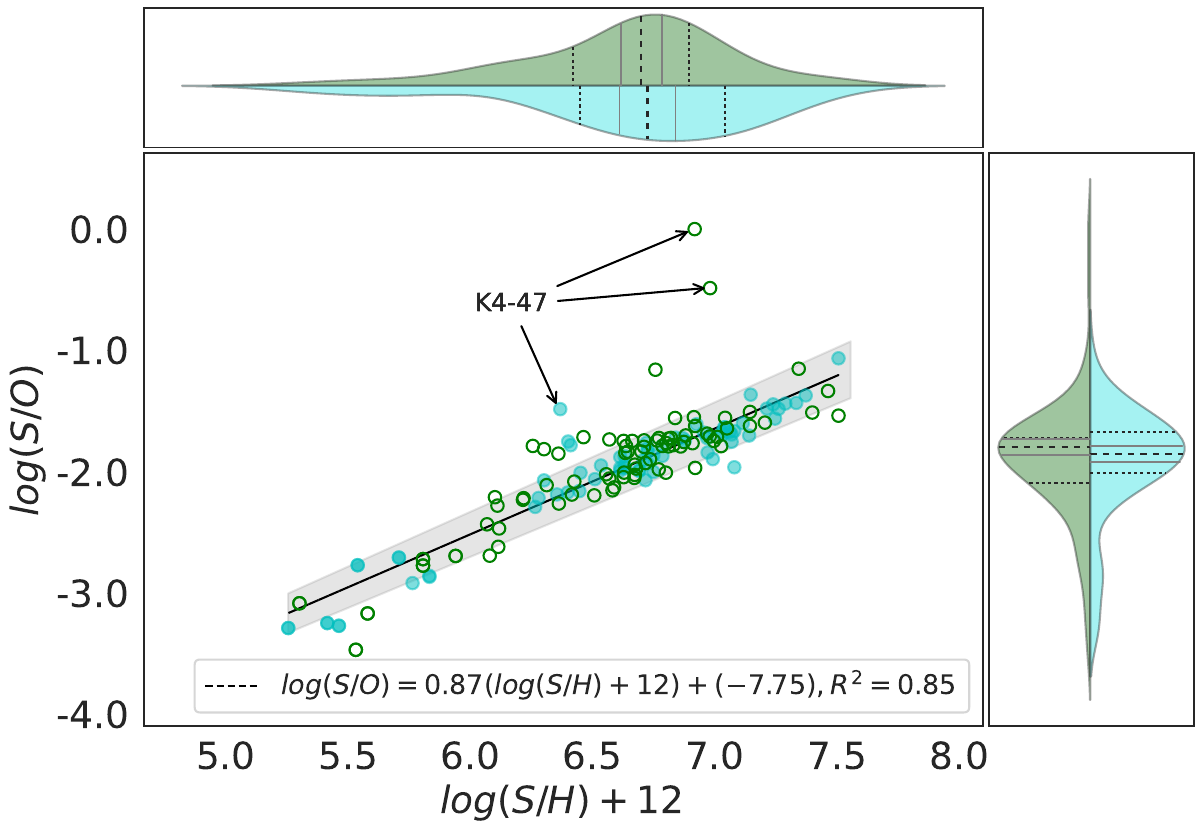}
    \includegraphics[width=0.498\textwidth]{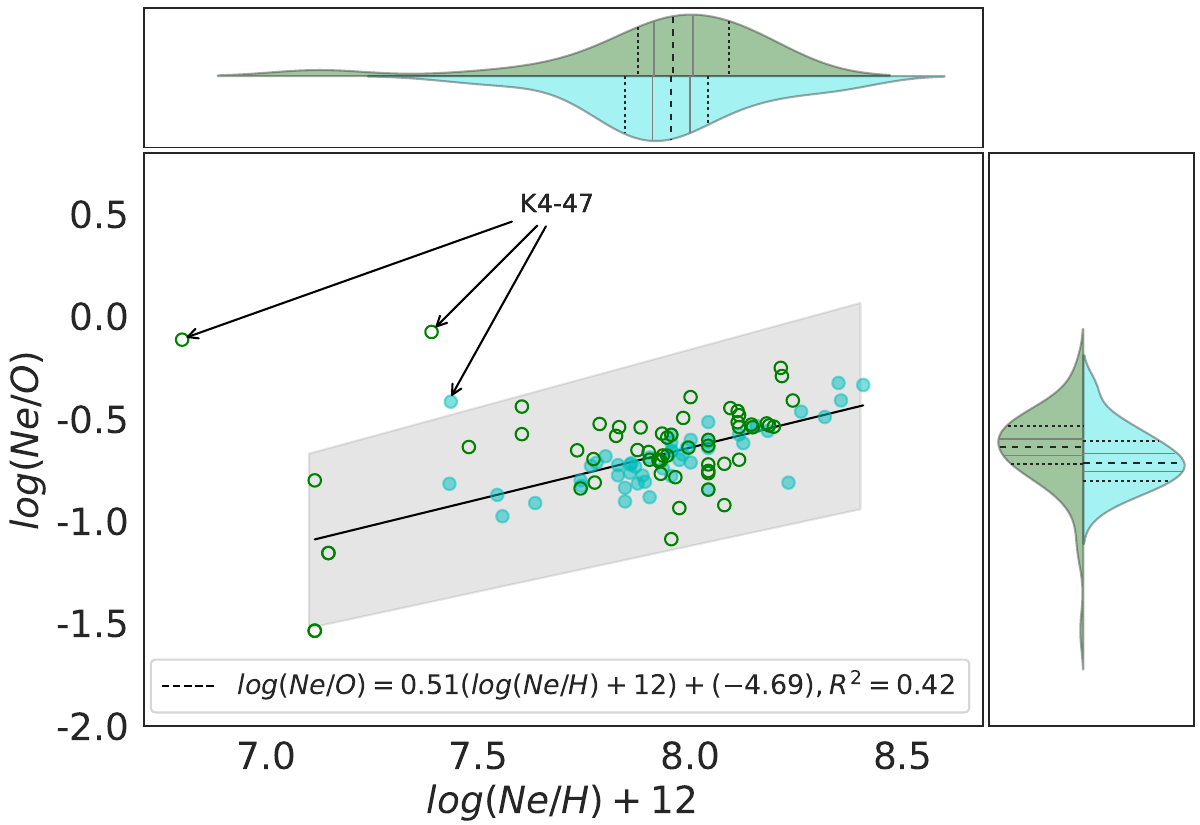}
    \includegraphics[width=0.498\textwidth]{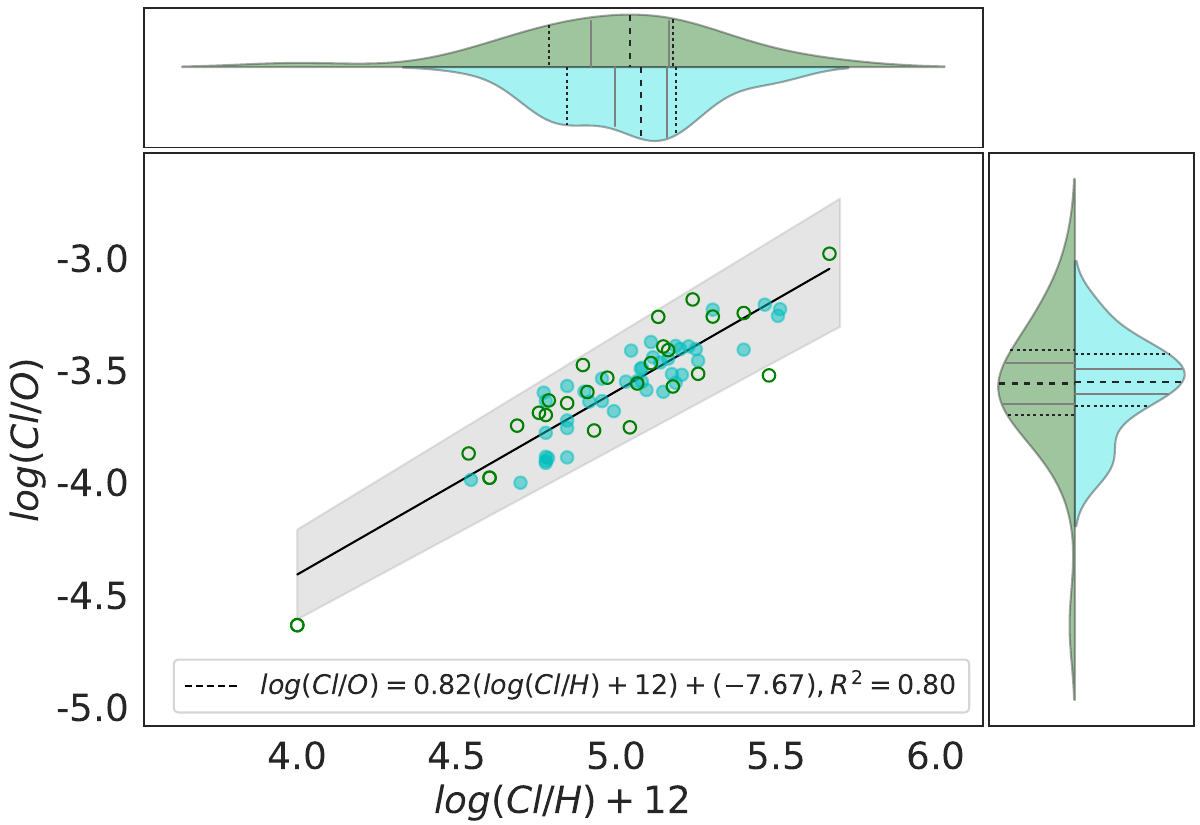}
    \caption{Correlations between total abundances using both components in the sample (Rims/Shells and LISs). Below each plot, the trend line with its goodness-of-fit (R$^2$) 
    per pair of elements is listed. The gray filled area represents the uncertainty of the regression line. Above and to the right of each panel, the split violinplots comparing Rims/Shells and LISs groups. }
    \label{fig:correl1}
\end{figure*}

Regarding the total abundance of N, Ar, S, Ne and Cl, in Fig.~\ref{fig:correl1} we present  different abundance ratio combinations \citep[see also, ][]{1990ApJ...356..229H}. From these correlations we do not find any specific locus for LISs or Rims/Shells which would indicate chemical inhomogeneities, but some structures notably distant from the linear correlation -- marked with solid arrows -- are identified. A particular example is K~4-47, which is composed of a compact high-ionization core and a pair of LISs with strong emission from low-ionization species, both studied by \citet{Goncalves2004}. A strong H$_2$ emission is associated to the LISs \citep{Akras2017}. Relative to the rest of PNe and LISs in the sample, this nebula shows significantly higher N/O, S/O and Ne/O abundance ratios. A second example is the Rims/Shells in Wray~17-1 \citep[][ named \textit{inner NEBs}]{Akras2016} for which a higher Ar/O ratio is found. A third example is the knots (or LISs) in the K~1-2 \citep[][]{Akras2016} for which a higher Ar/O ratio is also found. For this last PN, it has been found that the central star is a post-CE close binary system \citep{Exter2003}. 
Note that these three PNe display a highly collimated jet-like structure with knots present at the end of them \citep[][]{1996A&A...313..913C,Goncalves2004}.

\begin{table*}
\caption{Same as Table~\ref{tab:violinplot}, but for abundances. }
\label{tab:violinplot_abun}
\begin{tabular}{l|cccccccc}
\hline
            & \multicolumn{2}{c}{log(He/H)} & \multicolumn{2}{c}{log(O/H)+12}      & \multicolumn{2}{|c}{log(N/O)}     & \multicolumn{2}{c}{log(N/H)+12}  \\
               
\hline                                                                                                                                                   
            & Rims/Shells   & LISs          &  Rims/Shells & LISs                  & Rims/Shells   & LISs              & Rims/Shells   & LISs              \\
Mean        & -0.9741       & -0.9876       &  8.5421      & 8.5309                & -0.5625       & -0.4335           & 8.0387        & 8.1322            \\
Median      & -0.9666       & -0.9706       &  8.6021      & 8.5821                & -0.5301       & -0.4096           & 8.0934        & 8.2041            \\
Lower notch & -0.9824       & -0.9884       &  8.5657      & 8.5412                & -0.6224       & -0.4749           & 8.0209        & 8.1329            \\
Upper notch & -0.9507       & -0.9528       &  8.6384      & 8.6229                & -0.4377       & -0.3443           & 8.1660        & 8.2753            \\
IQR         &  0.0851       &  0.1015       &  0.1950      & 0.2326                &  0.4814       &  0.3697           & 0.3782        & 0.4030            \\
Q1          & -1.0132       & -1.0223       &  8.5044      & 8.4594                & -0.7583       & -0.6324           & 7.9015        & 7.9380            \\
Q3          & -0.9281       & -0.9208       &  8.6994      & 8.6920                & -0.2768       & -0.2627           & 8.2796        & 8.3410            \\
\#Outliers  & 0             &  3            &  8           & 6                     &  4            &  4                & 6             & 7                 \\
\#Sample    & 71            &  80           &  71          & 80                    &  67           &  79               & 67            & 79                \\
\hline                                                                                                                                                                 
            & \multicolumn{2}{c}{log(Ar/O)} & \multicolumn{2}{c}{log(Ar/H)+12} & \multicolumn{2}{c}{log(S/O)} & \multicolumn{2}{|c}{log(S/H)+12}           \\
\hline                                                                                                                                                    
            & Rims/Shells   & LISs          & Rims/Shells   & LISs             & Rims/Shells   & LISs         & Rims/Shells   & LISs                       \\
Mean        & -2.6381       & -2.3811       & 5.9654        & 6.2245           & -1.9413       & -1.9516      & 6.6829        & 6.6241                     \\
Median      & -2.4819       & -2.3440       & 6.1139        & 6.2788           & -1.8646       & -1.8129      & 6.7243        & 6.6776                     \\ 
Lower notch & -2.6244       & -2.4012       & 5.9717        & 6.2073           & -1.9291       & -1.8826      & 6.6092        & 6.5957                     \\ 
Upper notch & -2.3393       & -2.2867       & 6.2562        & 6.3502           & -1.8001       & -1.7433      & 6.8394        & 6.7595                     \\ 
IQR         &  0.7149       &  0.2552       & 0.7132        & 0.3188           &  0.3314       &  0.3893      & 0.5911        & 0.4578                     \\ 
Q1          & -3.0008       & -2.4907       & 5.6037        & 6.0792           & -2.0076       & -2.1099      & 6.4502        & 6.4150                     \\ 
Q3          & -2.2859       & -2.2355       & 6.3169        & 6.3979           & -1.6762       & -1.7205      & 7.0414        & 6.8727                     \\ 
\#Outliers  & 0             & 1             & 0             & 1                & 9             &  6           & 4             & 3                          \\ 
\#Sample    & 62            & 49            & 62            & 49               & 65            &  77          & 65            & 77                         \\ 
\hline

            & \multicolumn{2}{c}{log(Ne/O)} & \multicolumn{2}{c}{log(Ne/H)+12} & \multicolumn{2}{c}{log(Cl/O)} & \multicolumn{2}{c}{log(Cl/H)+12}         \\
\hline                                                                                                               
            & Rims/Shells   & LISs          & Rims/Shells   & LISs             & Rims/Shells   & LISs          & Rims/Shells   & LISs                     \\                                                                                                         
Mean        & -0.6894       & -0.6548       & 7.9519        & 7.9248           & -3.5712       & -3.5761       & 5.0412        & 4.9944                   \\      
Median      & -0.7129       & -0.6383       & 7.9542        & 7.9590           & -3.5504       & -3.5588       & 5.0755        & 5.0414                   \\      
Lower notch & -0.7565       & -0.6785       & 7.9098        & 7.9134           & -3.6062       & -3.6494       & 4.9942        & 4.9194                   \\      
Upper notch & -0.6693       & -0.5980       & 7.9987        & 8.0046           & -3.4946       & -3.4682       & 5.1569        & 5.1634                   \\      
IQR         &  0.1923       &  0.1884       & 0.1963        & 0.2135           &  0.2332       &  0.2886       & 0.3396        & 0.3886                   \\      
Q1          & -0.8017       & -0.7212       & 7.8451        & 7.8764           & -3.6607       & -3.6990       & 4.8451        & 4.7875                   \\      
Q3          & -0.6094       & -0.5328       & 8.0414        & 8.0899           & -3.4275       & -3.4103       & 5.1847        & 5.1761                   \\      
\#Outliers  &  0            &  3            & 5             & 4                & 0             & 1             & 0             & 1                        \\      
\#Sample    &  48           &  54           & 48            & 54               & 43            & 25            & 43            & 25                       \\

\hline

\end{tabular} 
\end{table*}

Fig.~\ref{fig:correl1} also includes the linear fit of the correlations, with the gray filled area corresponding to the uncertainties. 
The structures that deviated significantly from bulk, marked with solid arrows, were excluded from the linear fitting. In Table~\ref{tab:correl} we list the slope (\textit{a}), intercept (\textit{b}), goodness-of-fit (R$^{2}$) and the number of the data points considered, 
without 
LISs. These structures were eliminated with the intention to look for potential deviations due to this specific group. In order to compare the distributions between the 
two groups of PN structures, the top and right side of each of the panels in Fig.~\ref{fig:correl1} show the split violinplots of the abundance ratio correlations. In general, there are no major differences between the distributions 
nor between their median values (see Table~\ref{tab:violinplot_abun}). Moreover, also taking into account the upper and lower notches of the correlations, we conclude that LISs, rims and shells are similar in terms of total elemental abundances. There are a few 
structures that are outliers from both violinplots of each panel -- that were not marked to avoid confusion. 
Although the latter outliers correspond to values that 
deviate significantly from the distributions of each group, it can be observed that, in general, they do not deviate from the linear correlation. 
K~4-47, for which the total abundances of He, O, N, Ne and S were studied by \citet{Goncalves2004} is the only PN that is outlier for all the abundances correlations. As pointed out by the authors, neither the core nor the LISs of this PN can be explained by pure photoionization, and therefore its abundance ratios are not reliable.

For the widely studied log(N/H) versus log(N/O) diagram, we determine $log(N/O)=(0.73\pm0.03)\times[12+log(N/H)]-(6.38\pm0.23)$, R$^{2}=0.82$, very close to the previous result reported by \citet{Akras2016} ($log(N/O)=0.74\times[12+log(N/H)]-6.50$; with R$^{2}=0.88$) considering only five PNe though. \citet{2013A&A...558A.122G} also determined practically the same relation ($log(N/O)=0.73\times[12+log(N/H)]-6.50$; with R$^{2}=0.86$) but for a sample of PNe with [WR] and \textit{wels} CSPNe. According to \citet{2013A&A...558A.122G}, this linear relationship indicates that N-enrichment in PNe occurs independently of the O abundance,  being mainly due to the CN-cycle -- where N increases at the expense of C -- and not to the ON-cycle. This last one would reduce the O/H ratio in low-metallicity PNe with progenitor stars higher than $\sim 2 M_{\odot}$, which can be observed for $log(O/H)+12 \leq 8$ \citep[][]{2017RMxAA..53..151M}. In the upper left panel of Fig.~\ref{fig:correl1} it can be seen that this value is achieved for three nebulae corresponding to K~4-47, K~1-2 and Wray~17-1. This idea is reinforced just for one of those PNe by looking at the upper right panel, in which the log(N/O) is lower than 0.5 for all 
structures, except the LISs in K~4-47. According to \citet{2010RMxAA..46..159C,2017MNRAS.468..272C}, PNe with log(N/O)$\sim$0.5 could be originated from massive stars, i.e., the higher N/O ratio, the more massive progenitor stars. 

The result from this work is, considering the errors, similar to the previously published ones. Nevertheless, for comparison purposes, in Table~\ref{tab:correl} we list the parameters of the linear fitting, 
excluding the LISs. Globally, the correlation with and without 
LISs do not vary strongly -- variations are at most of $\sim$6\% on slope and intercept -- as highlighted by the R$^{2}$ values, except for Ne/O. Remembering that R$^{2}$ ranges from 0 to 1 --~the higher the value, the better the fit~-- the table shows that 
without the LISs, the coefficients are closer to 1. 

\begin{table}
\centering
\caption{Slope, intercept, goodness-of-fit and number of the sample of the fitted trend lines for correlations between different abundance ratios without the LISs. In parentheses, the values for the correlations using the two groups (Rims/Shells and LISs) are specified.}
\label{tab:correl}
\begin{tabular}{l|cccc}
\hline
                            & \textit{a}          & \textit{b}          & R$^{2}$ & \#Sample \\
\hline
\multirow{2}{*}{log(N/O)}   & 0.71$\pm$0.03       & -6.28$\pm$0.27      & 0.88    & 67        \\
                            & (0.73$\pm$0.03)     & (-6.38$\pm$0.23)    & (0.82)  & (146)     \\
\hline                            
\multirow{2}{*}{log(Ar/O)} & 0.93$\pm$0.06        & -8.17$\pm$0.37      & 0.79    & 62        \\
                           & (0.88$\pm$0.05)      & (-7.90$\pm$0.28)    & (0.77)  & (111)      \\
\hline                            
\multirow{2}{*}{log(S/O)}  & 0.88$\pm$0.03        & -7.81$\pm$0.23      & 0.91    & 65        \\
                           & (0.87$\pm$0.03)      & (-7.75$\pm$0.20)    & (0.85)  & (142)     \\
\hline                            
\multirow{2}{*}{log(Ne/O)} & 0.57$\pm$0.06        & -5.21$\pm$0.48      & 0.66    & 48        \\
                           & (0.51$\pm$0.06)      & (-4.69$\pm$0.48)    & (0.42)  & (102)     \\
\hline                            
\multirow{2}{*}{log(Cl/O)} & 0.81$\pm$0.06        & -7.65$\pm$0.32      & 0.80    & 43        \\
                           & (0.82$\pm$0.05)      & (-7.67$\pm$0.25)    & (0.80)  & (68)      \\ 
\hline                           
\end{tabular}

\end{table}

Altogether, these abundance ratio correlations are suggesting that Ar, S, Ne and Cl vary in lockstep with O, which means that the former element's nucleosynthesis during the evolution of the progenitor star, as compared to the latter, are small -- or even negligible \citep{2010RMxAA..46..159C,2017RMxAA..53..151M}.

\subsection{Excitation mechanism} \label{sec_exc}

Aiming to understand the nature of the nebula S176, \citet*{Sabbadin1977} (hereafter SMB) introduced a diagnostic diagram (DD), 
based on the \ha/\nitrogen~$\lambda\lambda$6548,6584 and \ha/\sulfurt~$\lambda\lambda$6716,6731 
line ratios, to distinguish PNe from H{\sc ii} regions and supernova remnants (SNRs). The position of these 
nebulae is distinctive because of the excitation mechanism and physico-chemical properties responsible for the production of the emission-line involved. 
\citet{2006RMxAA..42...47R} used a larger sample of PNe and provided a more restrict region through a density probability ellipse, on the same DD \citep[see also][]{Frew2010,2013MNRAS.431..279S,akras2020a}.

\begin{figure}
    \centering
    \includegraphics[width=\columnwidth]{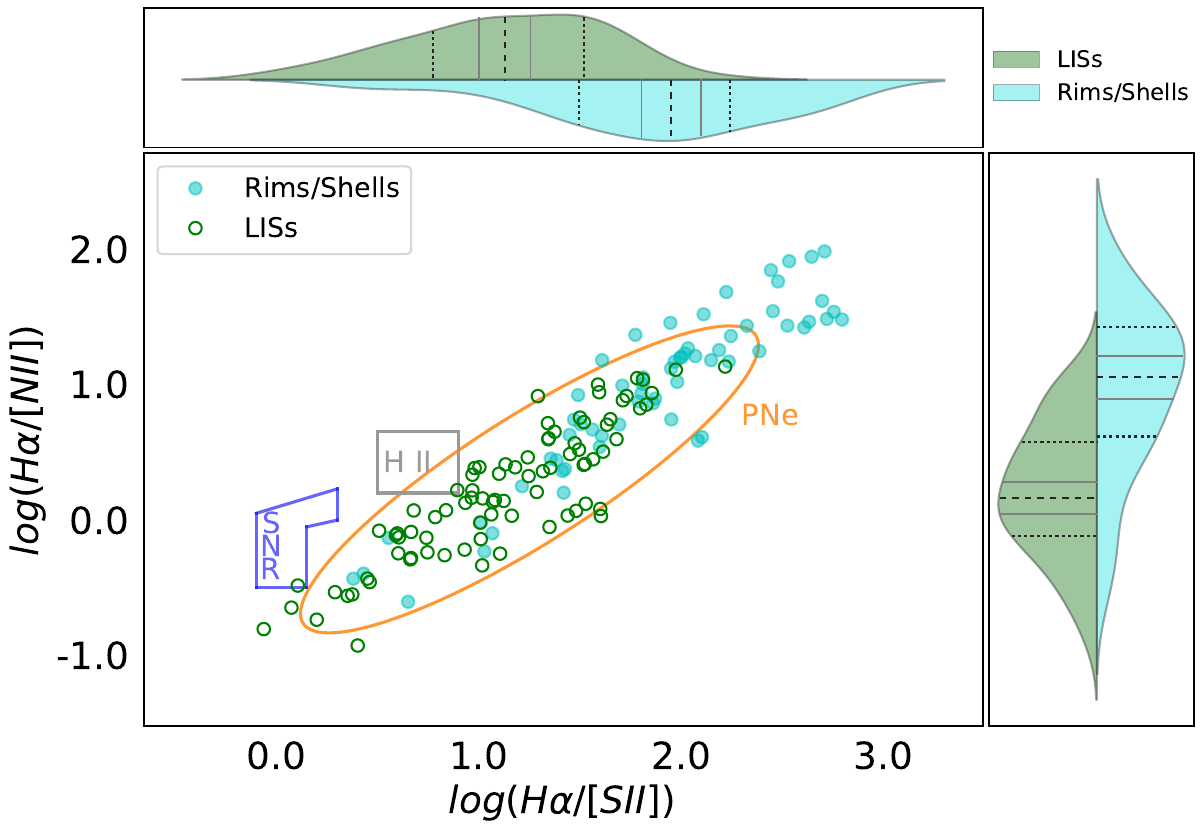}
    \caption{\ha/\nitrogen~6548+6584 versus \ha/\sulfurt~6716+6731 diagnostic diagram \citep{Sabbadin1977}  with the density ellipse of probability 0.85 from \citet{2006RMxAA..42...47R}.}
    \label{fig:sab}
\end{figure}

In Fig.~\ref{fig:sab} we display SMB for our sample of PNe, Rims/Shells and LISs\footnote{It is important to note that the diagrams under discussion were developed using the integrated emission of many PNe. Here, as first proposed by \citet{Goncalves2003}, they are used for different components of the same PN, in a spatially resolved fashion. The differences of integrated and spatially resolved analysis are also discussed in \cite{akras2020a,Akras2022}}. The disparity in the \ha/\nitrogen~and \ha/\sulfurt~line ratios between 
LISs and Rims/Shells is evident in this DD. Most LISs, by definition characterized by enhanced \nitrogen~ and \sulfurt~lines relative to \ha,  lie in the bottom-left half of the density ellipse with log(\ha/\nitrogen)$<$1.0 and log(\ha/\sulfurt)$<$1.8. On the other hand, the vast majority of the Rims/Shells 
are distributed in the top-right of the PNe locus in the diagram. 
This becomes even more evident when looking at the split violinplots corresponding to each of the axes. Taking into account the upper and lower notches 
(see Table~\ref{tab:vio1}) it is evident that the two groups are statistically different. 
This separation between LISs and Rims/Shells is attributed to the difference in the ionization state 
of these nebular components \citep{Akras2020b}, which can also be interpreted as excitation stratification 
\citep[][]{Goncalves2003}.

\begin{figure}
    \centering
    \includegraphics[width=\columnwidth]{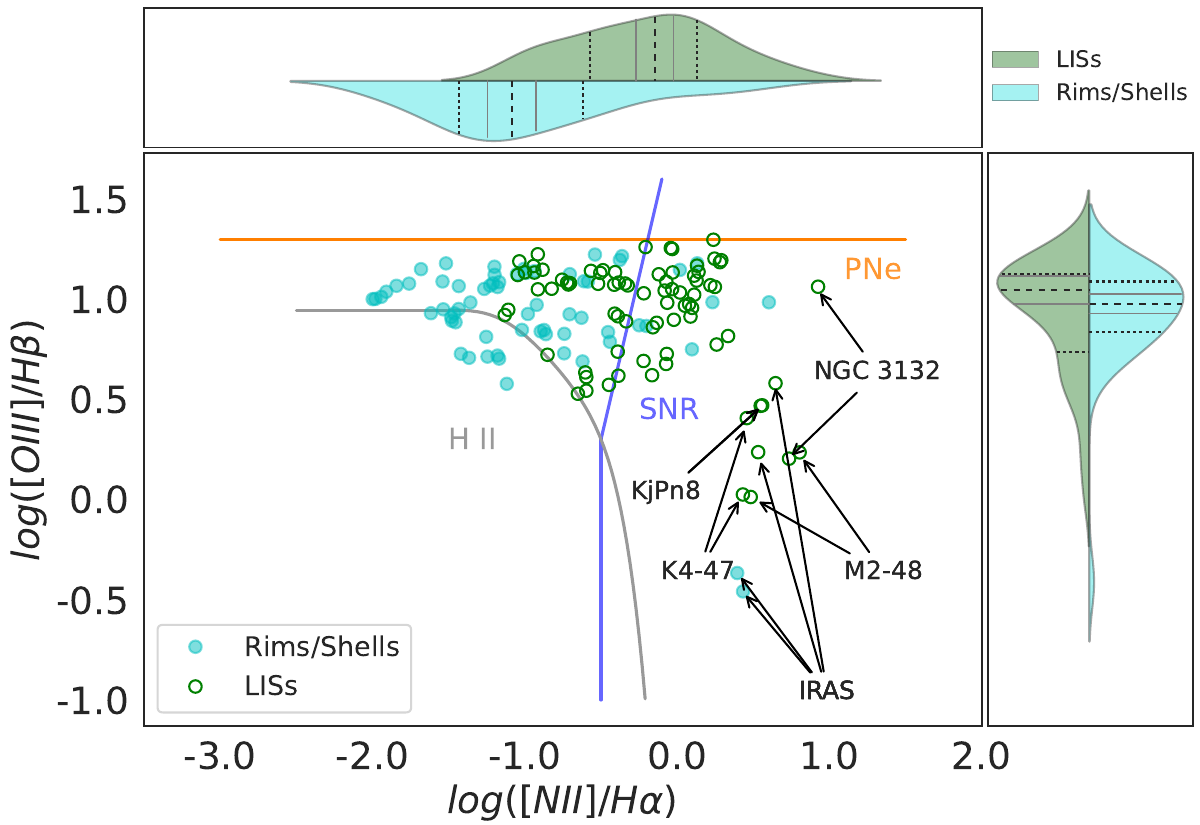}
    \includegraphics[width=\columnwidth]{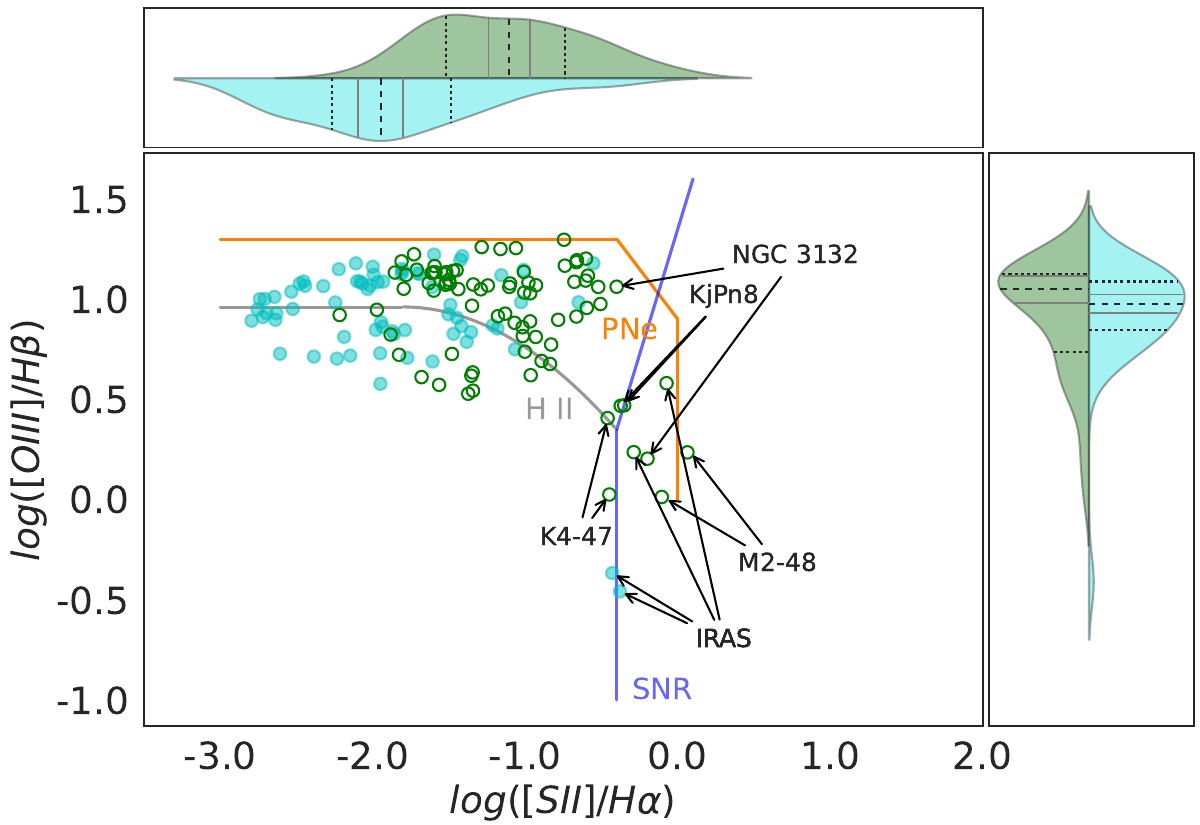}
    \caption{\textit{Top panel:} \nitrogen~6548+6584/\ha~versus \oxygeniii~5007/\hb~BPT diagnostic diagram with the regions of PNe, H~II regions and SNRs \citep{Frew2010,2013MNRAS.431..279S}. \textit{Lower panel:} \sulfurt~6716+6731/\ha~versus \oxygeniii~5007/\hb~BPT diagram.} 
    \label{fig:bpt}
\end{figure}

Two other DDs were proposed by \citet*{1981PASP...93....5B} (hereafter BPT) to explore the excitation mechanisms in galaxies. The corresponding BPT diagrams for PNe, H{\sc ii} regions and SNRs were discussed by \citet[][]{Frew2010}. Fig.~\ref{fig:bpt} shows the distributions of LISs and Rims/Shells in the BPT diagrams, together with their violinplots. Some particular cases that deviate from the bulk of the data with apparent enhanced \sulfurt/\ha~and \nitrogen/\ha~ratios are indicated with arrows.
Two cyan data points (Rims/Shells) exhibit the lowest \oxygeniii/\hb~ratio and correspond to the lobes of the water maser emitting PN (H$_{2}$O-PN) IRAS~18061–2505 \citep{2021arXiv210505186M}, while its \sulfurt/\ha~and \nitrogen/\ha~ratios are significantly high. Two LISs of this young nebula are described by stronger \oxygeniii~$\lambda$5007 emission and comparable \sulfurt/\ha~and \nitrogen/\ha~ratios. \citet[][]{2021arXiv210505186M} argued that the optical spectra of the bow-shock structures (LISs for our definition) of H$_{2}$O-PN are attributed to shock interactions, while the photoionization dominates the spectra of the lobes (Rims/Shells \textbf{here}), if the mass of the progenitor star is $\gtrsim3M_{\odot}$. Two LISs with high \sulfurt/\ha~and \nitrogen/\ha~ratios are also found in KjPn~8 \citep{2009MNRAS.398.2166G}. These authors concluded that both LISs are consistent with both shock- and photoionization playing a role in the measured emission line ratios.

Another two LISs, from the unusual PN K~4-47 \citep{Goncalves2004}, are also characterized by high \sulfurt/\ha~and \nitrogen/\ha~ratios. Based on the prediction from shock modelling, both LISs are shock dominated. The high H$_2$~(1-0)/(2-1) ratio measured for these structures is also attributed to shock-heated gas \citep[][]{2001MNRAS.328..419L,Akras2017}, but the hypothesis of a high density gas (>10$^{4}$~cm$^{-3}$) illuminated by an intense UV radiation field has not been ruled out yet. 
Two more LISs, from M~2-48, exhibit high \nitrogen/\ha~and \sulfurt/\ha~ratios. This nebula with its multiple knots (LISs) was studied by \citep{2002AA...388..652L}, who conclude that shock
excitation is contributing to the spectra of both LISs analysed in contrast to the central region, which is radiatively excited.
Finally, the two arc-like structures found in NGC~3132 \citep{2020A.A...634A..47M} also exhibit high \sulfurt/\ha~and \nitrogen/\ha~ratios. IRAC images from {\it Spitzer} have revealed mid-IR emission at the position of these arc-like structures, likely from H$_2$ lines \citep[][]{2004ApJS..154..296H}. Early release images from JWST confirmed the presence of H$_2$ emission throughout this nebula \citep{2022NatAs...6.1421D}. Although, it is likely that unidentified infrared emission bands (UIBs) detected in NGC~3132 \citep[]{Mata2016} also contribute to the {\it Spitzer's} images.  

From the statistical point of view, LISs and Rims/Shells subsets are different groups in terms of the \nitrogen/\ha~and \sulfurt/\ha~line ratios. The aforementioned LISs are characterized by lower \oxygeniii/\hb~ratios ($\lesssim$0.5) relative to the main bulk of data points (see Table~\ref{tab:vio1}) and significantly higher \nitrogen/\ha~and \sulfurt/\ha~line ratios. We thus argue that the shock-heating process is likely active in these particular cases, which are prone to further studies. Note that all these LISs exhibit \sulfurt/\ha$>$0.4, a strong tracer of shocks \citep[e.g.][]{Leonidaki2013,kopsacheili2020}. The shock velocity in these cases should be $\leq$100~km~s$^{-1}$.

\begin{table*}
\caption{Group properties for the samples of Rims/Shells and LISs for log(H$\alpha$/[S~{\sc ii}]6717+6731), log(H$\alpha$/[N~{\sc ii}]6548+6584), log([O~{\sc iii}]5007/H$\beta$), log([O~{\sc i}]6300/H$\alpha$ and log([O~{\sc ii}]3727/[O~{\sc iii}]5007). The notches correspond to the approximated 95 per cent CIs \textbf{(see Appendix~\ref{apA})}. The last two rows show the number of outliers and sample size, respectively.}
\label{tab:vio1}
\centering
\begin{tabular}{l|cccccccccc}
\hline
                 &  \multicolumn{2}{c}{log(H$\alpha$/[S~{\sc ii}])} &  \multicolumn{2}{c}{log(H$\alpha$/[N~{\sc ii}])} &  \multicolumn{2}{c}{log([O~{\sc iii}]/H$\beta$)}                                                                        &  \multicolumn{2}{c}{log([O~{\sc i}]/H$\alpha$)} &  \multicolumn{2}{c}{log([O~{\sc ii}]/[O~{\sc iii}])} \\
\hline                                                                                                                
               &  Rims/Shells  &  LISs     &  Rims/Shells  &    LISs      &     Rims/Shells  &    LISs         &  Rims/Shells  &    LISs    &  Rims/Shells  &    LISs     \\
Mean           & 1.8680        &  1.1248   &  0.9510       &   0.2122     &     0.9221       &    0.9163       &  -2.2603      &  -1.3055   &  -1.3525      &   -0.7563   \\
Median         & 1.9501        &  1.1271   &  1.0517       &   0.1623     &     0.9718       &    1.0461       &  -1.9931      &  -1.1894   &  -1.3858      &   -0.7311   \\
Lower notch    & 1.8031        &  0.9997   &  0.8915       &   0.0443     &     0.9222       &    0.9775       &  -2.3557      &  -1.3366   &  -1.5326      &   -0.8528   \\
Upper notch    & 2.0971        &  1.2544   &  1.2120       &   0.2803     &     1.0214       &    1.1146       &  -1.6305      &  -1.0421   &  -1.2389      &   -0.6095   \\
IQR            & 0.7433        &  0.7436   &  0.8102       &   0.6889     &     0.2507       &    0.3906       &   1.1315      &   0.6221   &   0.4582      &    0.5139   \\
Q1             & 1.4969        &  0.7746   &  0.6165       &  -0.1154     &     0.8376       &    0.7354       &  -2.9063      &  -1.5782   &  -1.5733      &   -0.9883   \\
Q3             & 2.2402        &  1.5182   &  1.4268       &   0.5735     &     1.0883       &    1.1259       &  -1.7748      &  -0.9560   &  -1.1151      &   -0.4744   \\
\#Outliers     &  1            &  0        &  1            &    0         &     2            &    2            &  0            &  2         &  0            &    0        \\
\#Sample       &  63           &  84       &  63           &    84        &     64           &    80           &  24           &  44        &  24           &    44       \\

\hline
\end{tabular} 
\end{table*}

\begin{figure}
    \centering
    \includegraphics[width=\columnwidth]{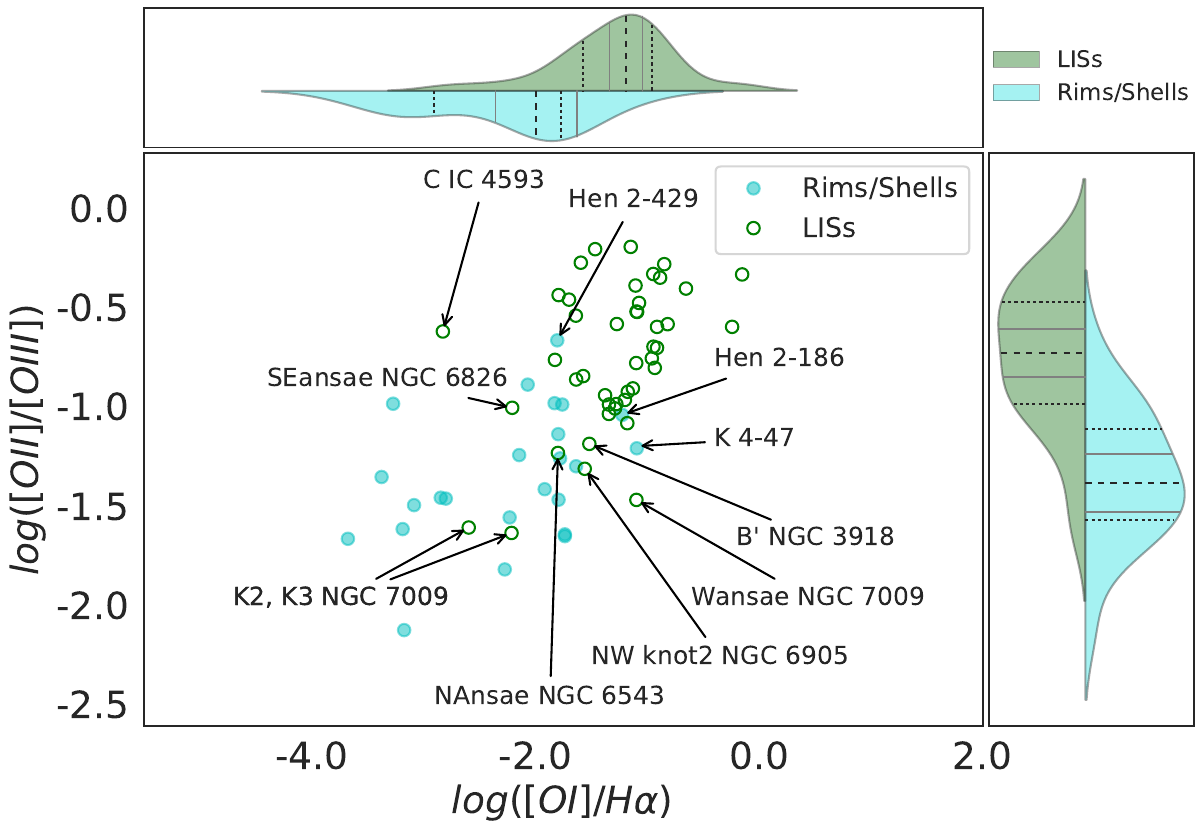}
    \includegraphics[width=\columnwidth]{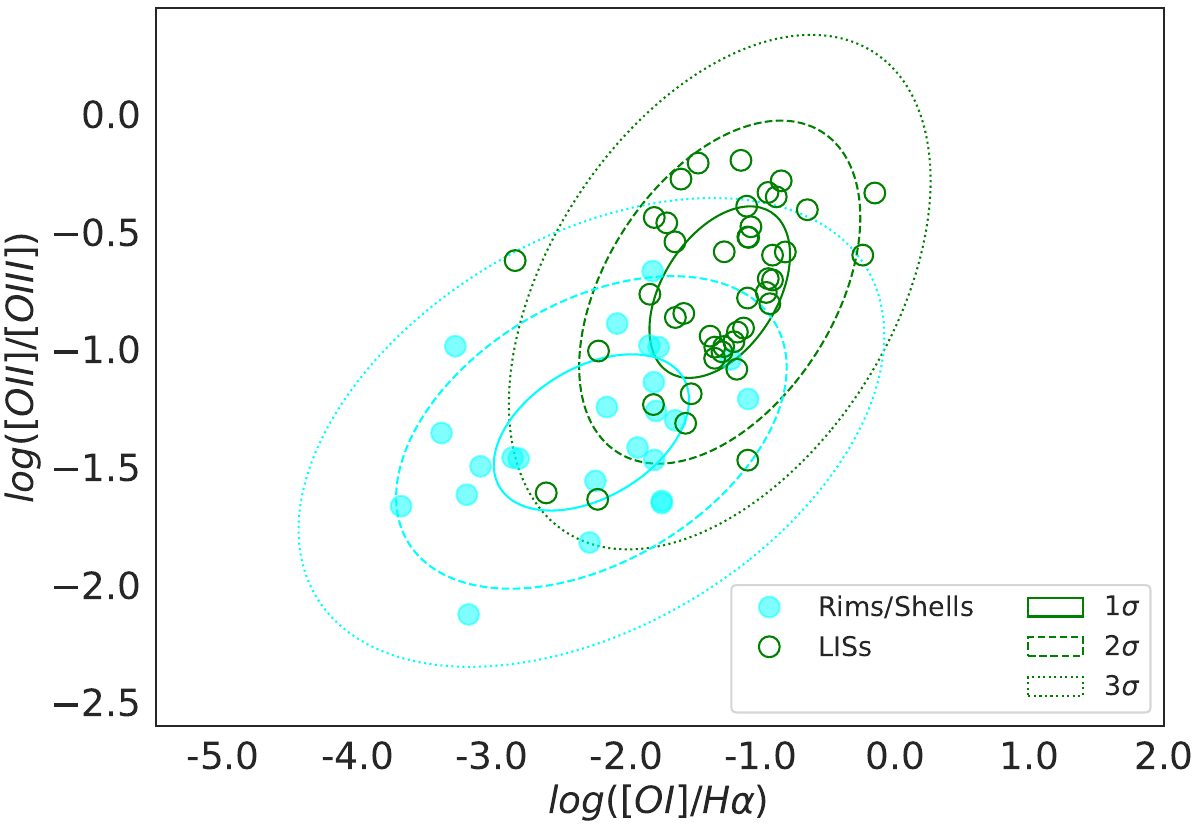}
    \caption{\textit{Upper panel:} log([O~{\sc i}]6300/H$\alpha$) versus log([O~{\sc ii}]3727 /[O~{\sc iii}]5007) diagram, with some 
    structures --~either LISs or Rims/Shells~ -- that move away from the bulk of the data in their respective group lying in the transition zone marked with arrows (see the text). \textit{Lower panel:} Same diagram, with the confidence ellipses of 1, 2 and 3$\sigma$.} 
    \label{fig:dds}
\end{figure}

To further explore the ionization state of the PNe in our sample, we also built the \oxygenii~$\lambda$3727/\oxygeniii~$\lambda$5007 versus \oxygeni~$\lambda$6300/\ha~diagram for both LISs and Rims/Shells subsets (Fig.~\ref{fig:dds}). 
The advantage of this DD is that only one chemical element in used, unlike the \textbf{rest} DDs.
A clear separation between LISs and Rims/Shells is observed. The former occupy the top-right corner in the plot, with high \oxygenii~$\lambda$3727/\oxygeniii~$\lambda$5007 and \oxygeni~$\lambda$6300/\ha~line ratios, while 
Rims/Shells are found to be concentrated in the bottom-left corner with lower line ratios. 
The bottom panel in Fig.~\ref{fig:dds} illustrates the same plot 
with the confidence ellipses of 1, 2, and 3$\sigma$. Note that the two groups display different slopes, which indicate an important alteration of the ionization state between the two components. Their violinplots and statistical parameters (Table~\ref{tab:vio1}) make it clear that LISs and Rims/Shells are certainly different samples. Nonetheless, it is important to note that, because the \oxygeni~$\lambda$6300 is not always detectable, the number of data points in Fig.~\ref{fig:dds} is lower than in the previous DDs -- having 44 LISs and 24 Rims/Shells -- so the results should be treated with caution. 

One can see that there are LISs 
into the locus of Rims/Shells and vice versa. This could be interpreted as a transition zone. 
Particularly, three LISs in NGC~7009 that corresponding to the K2 and K3 knots \citep{Goncalves2003} and western ansae \citep{1994ApJ...424..800B} have lower \oxygenii~$\lambda$3727/\oxygeniii~$\lambda$5007 ratio relative to the bulk of LISs. A similar behaviour is also observed for the northern ansae in NGC~6543 \citep{1994ApJ...424..800B}, the northwest knot2 in NGC~6905 and the B' LIS in NGC~3918, contrary to the C-LIS in IC~4593 \citep{2023MNRAS.518.3908M} and southeast ansae NGC~6826 which are characterized by low \oxygeni$\lambda$6300/\ha~ratio and high  \oxygenii~$\lambda$3727/\oxygeniii~$\lambda$5007.
On the other hand, there are three cases of Rims/Shells (corresponding to Hen~2-186, Hen~2-429 and K~4-47) with high \oxygenii~$\lambda$3727/\oxygeniii~$\lambda$5007 and \oxygeni$\lambda$6300/\ha~ratios placed into the regime of LISs. Hen 2-186 is a poorly studied southern PN and, according to \citet{Guerrero2020b}, it belongs to a limited group of nebulae whose jets have velocities exceeding 100~kms$^{-1}$. Hen 2-429 is also a PN that belong in PNe with embedded jets and finally, K~4-47 with a collimated structure and a pair of shock-heated LISs. 
While these three last examples are Rims/Shells and do not include the jets, filaments or knots of low-ionization, 
they may also contain some shock excitation contribution. 
Taking into account that LISs and Rims/Shells occupy two separate regions with different slopes, we conclude that the ionization state of the nebular gas in the two groups is certainly different.

\section{Photoionization versus shock model predictions}
\label{sec4}

\begin{figure*}
\centering
\includegraphics[width=19.5cm]{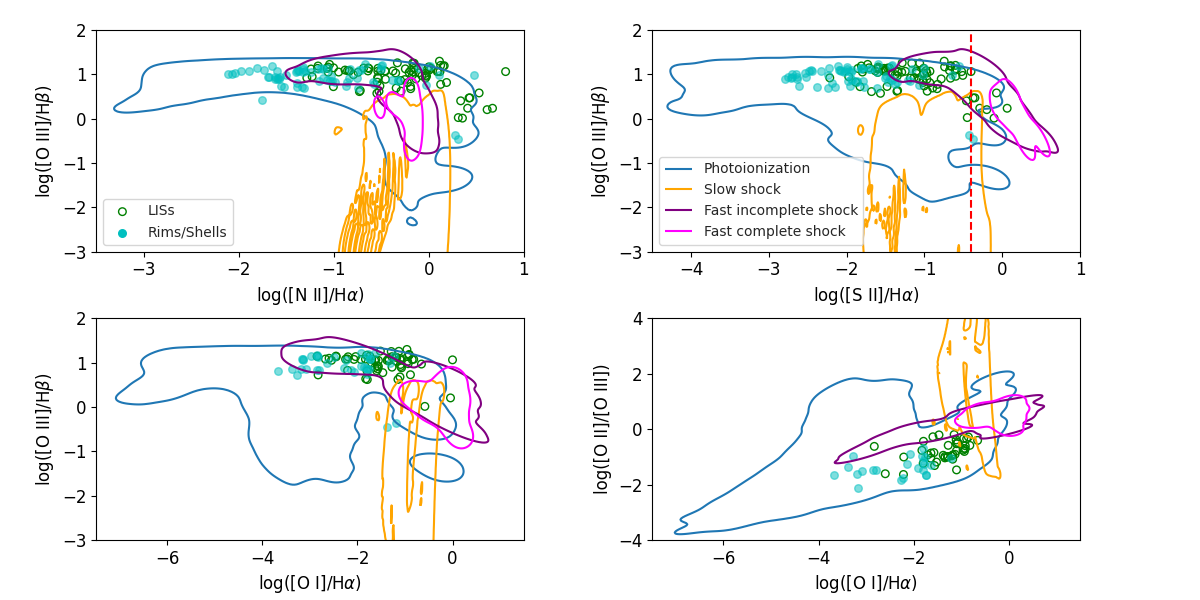}
\includegraphics[width=19.5cm]{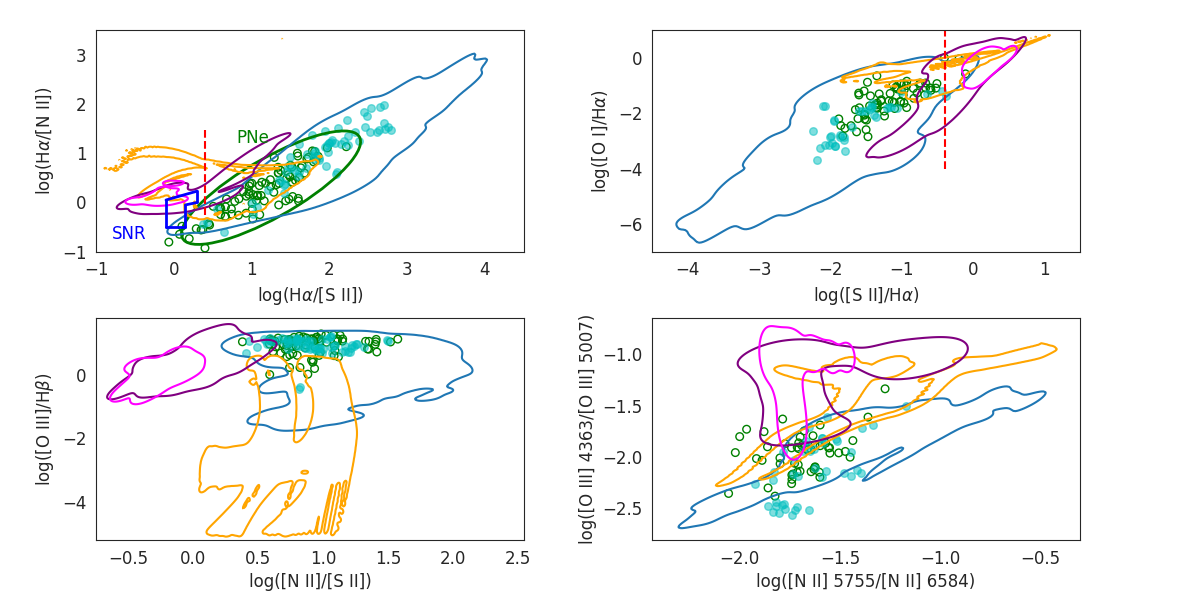}
\caption[]{Emission-line diagnostic diagrams generated from grids of photoionization and shocks models available in the 3MBD. The photoionization models are presented with a blue contour, the low-velocity with an orange contour and the complete/incomplete high velocity shocks models with purple/magenta contours, respectively. The different contours cover the 98~percent of the total grids. Our sample of LISs and Rims/Shells are shown with unfilled green circles and filled cyan circles. The common log(\sulfurt/\ha)$>$-0.4 selection criterion for shock-excited supernova remnants \protect\citep[e.g.][]{Leonidaki2013, kopsacheili2020} is shown with a dashed-red line.}
\label{UVShmodels}
\end{figure*}

The disentanglement of UV photo-heated and shock-heated gases in the SMB and BPT DDs (Figs~\ref{fig:sab} and \ref{fig:bpt}) is still not well-defined, as both mechanisms yield comparable line ratios. The availability of two large grids of photoionization and shock models in the Mexican Million Models database \citep[3MdB,][]{Morisset2015,Alarie2019} allows us to explore the range of line ratios for both excitation mechanisms and for a wide range of physical parameters (see Appendix~\ref{apB} for more details about the grid of models).

In Figure~\ref{UVShmodels}, we present a number of emission line DDs, including the common BPT and SMB diagrams, combining the predictions from the grids of photoionization and shock models. Only the photoionization models that satisfy the criteria in \citet{2014MNRAS.440..536D} with a sub-solar abundances set (log(O/H)=-3.66), a black-body approximation for the energy distribution of the central source and constant density law are used. On the other hand, the grid of shock models is constrained by the following properties. i) Cut-off temperature (T$_{\rm cut-off}$) and pre-shock temperature (T$_{\rm pre-shock}$) both, $<$12500~K. ii) Shock-velocity between 10 and 100~kms$^{-1}$ \citep[slow shock models,][]{Alarie2019} and between 100 and 350~kms$^{-1}$ \citep[fast-shock models,][]{Allen2008,Morisset2015}. iii) A sub-solar abundances set (namely, Allen2008\_Dopita2005) for both subsets. iv) A pre-shock density between 10 and 100~cm$^{-3}$ for the slow-shock models and 1~cm$^{-3}$ for the fast-shock models, for comparability reasons. And v) a transverse magnetic field $<$10$\mu$G for both subsets of shock models. 

An important overlap between the photoionization and shock models, especially for the low-ionization line ratios, is apparent. It is also evident that the common log(\ha/\sulfurt)$<$0.4 criterion for shock-excited supernova remnants \citep[e.g.,][]{Leonidaki2013,kopsacheili2020} does not adequately distinguish the two mechanisms. Although, it is possible to get such low \ha/\sulfurt~ratio from photoionization models with logU$<$-3.5 (see upper, right panel in Figure~\ref{UVShmodels}), resembling low- and high-velocity shock models.

Based on the BPT DDs, the low-velocity shock models (orange contours) cover a wide range of \oxygeniii/\hb~ values from -5 to 0.5 (in logarithmic scale), but narrower ranges are covered by the \oxygeni/\ha, \sulfurt/\ha~ and \nitrogen/\ha~line ratios (Fig.~\ref{UVShmodels}). On the other hand, the complete (see caption of Fig.~\ref{UVShmodels}) fast-shock models yield, \oxygeniii~$\lambda$5007/\hb~between -1 and 1. It should be noted that log(\oxygeniii~$\lambda$5007/\hb)>0.5 is only produced by fast-shock models and photoionization models. Therefore, the \oxygeniii~$\lambda$5007/\hb~ratio is a tracer of fast and slow shocks. Note that, shock models that produce log\oxygeniii/\hb$\sim$1 and \oxygeni/\ha$\sim$-2.0 are characterized by T$_{\rm cut-off}\gtrsim$12000~K (the higher the T$_{\rm cut-off}$ the lower the \oxygeni/\ha~ratio; see Appendix~\ref{apB}). The increase of the \oxygeni/\ha~and \sulfurt/\ha~line ratios in fast-shock models is followed by a decrease of the \oxygeniii/\hb~ratio, while the low-velocity shocks models 
do not show this dependency (see at the BPT diagrams).

The \oxygenii/\oxygeniii~versus \oxygeni/\ha~diagnostic diagram can also be used to determine the dominant excitation mechanisms of the nebulae. Fast shock (complete models) are restricted to a very narrow range of values (-0.5$<$log(\oxygenii/\oxygeniii)$<$1), while the slow shock models have a minimum value of $\sim$-2 and significantly higher values than fast shock models. The incomplete fast shock models can yield to lower \oxygenii/\oxygeniii~and \oxygeni/\ha~line ratios. On the other hand, most of the photoionization models display a linear increase of \oxygenii/\oxygeniii~as a function of the \oxygeni/\ha~ratio, as we can observe in Figure~\ref{UVShmodels}. There are though models that break this relation, with 0<log(\oxygenii/\oxygeniii)<2 and -4$<$log(\oxygeni/\ha)$<$-2, and they are characterized by low-temperature central sources ($<$60000~K; see Appendix~\ref{apB}).

The locus of PNe and supernova remnants (SNRs) in the SMB (\ha/\nitrogen~6548+6584 versus \ha/\sulfurt~6716+6731) diagnostic diagram are also displayed. The bulk of shock models do not totally coincide with the region of observed SNRs, and this is attributed to the chemical abundances of these models. The lower the abundance of N, lower the \nitrogen~6548+6584/\ha. The same result was also reported by \citet{Leonidaki2013} based on the spectroscopic observations of several SNRs in galaxies with different metallicities. The \ha/\sulfurt~6716+6731 ratio also follows the same relation with abundances. The lower the abundance of S, lower the \sulfurt~6716+6731/\ha\ ratio.

The \oxygeni/\ha~versus \sulfurt~6716+6731/\ha~ diagram shows a small overlap between the predicted line ratios of the two mechanisms, being among the most crucial diagnostics for disentangling the shock-heated and photo-heated nebulae, as it has already been shown \citep[e.g.][]{Phillips1998,Leonidaki2013,Akras2016,kopsacheili2020}. Scrutinizing the models from 3MdB, we found that only photoionization models with logU$<$-3 are able to produce 
line ratios compatible to those of shock models. Therefore, this diagram can be a very helpful diagnostic tool. We argue that a selection 
criteria log(\oxygeni/\ha)$>$-2 in conjunction with the common log(\sulfurt~6716+6731/\ha)$>$-0.4 can provide shock-heated nebulae with high confidence and fewer contaminants. 

The diagnostic diagram based on \oxygeniii/\hb~ versus \nitrogen/\sulfurt~line ratios  is also presented in Fig.~\ref{UVShmodels}. High-velocity shock models display a clear separation from the bulk of photoionization models, with \nitrogen/\sulfurt$<$0.4. On the contrary, the slow-velocity shock models overlap with photoionization models for 0.4$<$log(\nitrogen/\sulfurt)$<$0.9. Hence, the \nitrogen/\sulfurt line ratio can also be useful to constrain at least the shock velocity.

The last diagnostic diagram explored in this work involves the temperature sensitive line ratios, \oxygeniii~4363/5007 and \nitrogen~5755/6584, and it provides a better separation between the shock-heated and photoionized nebular gases. An upper bound in log(\oxygeniii~4363/5007) is found for the photoionization mechanism of $\sim$~-1.5, while there is no difference in log(\nitrogen~5755/6584). The overlap between shock and photoionization models is minimal, and a selection criterion log(\oxygeniii~4363/5007)$\geq$-1.5 can also be suitable for determining the excitation mechanism. Log(\oxygeniii~4363/5007) never takes values lower than -1.1 in shock models with transverse magnetic field <5$\mu$G. This agrees with the results from the bow-shock models of~K 4-47 with log(\oxygeniii~4363/5007)$\sim$-1, significantly higher than the observed values -1.3 and -1.5 of the two knots \citep{Goncalves2004}. The pre-shock magnetic field in these bow-shock models was considered negligible. Therefore, in case shocks take place in K~4-47 due to the highly moving knots, a magnetic field $>$5$\mu$G should be present. A few LISs are found to agree with low-velocity shock models characterized by log(\nitrogen~5755/6584)$\lesssim$-1.7 and log(\oxygeniii~4363/5007)$\geq$-2.2, but the UV photoionization process cannot be ruled out. 

According to the analysis above, we conclude that there is a significant overlap between the modelled predictions from the two mechanisms depending on the line ratio, and it is not feasible to disentangle the photo-heated and shock-heated gases based on individual line ratios, 
a combination of different line ratios is more efficient. The comparison of the observations with the models does not support the scenario in which shock interactions is the dominant mechanism for the majority of the LISs.

\section{Discussion}\label{sec5}

In order to obtain more reliable and comprehensive conclusions regarding the LISs in PNe, a statistical analysis of their physicochemical properties and emission line ratios was carried out using the largest sample gathered so far. In the following, we discuss the different aspects addressed throughout the work, trying to emphasize, if present, 
the variations between the LISs, rims and shells of PNe.

\subsection{Electron density} 
Considering a sample of 79 Rims/Shells and 98 LISs, the comparison between the two groups in Fig.~\ref{fig:violinplotTN} and Table~\ref{tab:violinplot} clearly shows that indeed LISs represent a statistically different sample than Rims/Shells in terms of electron density, as previously shown for several PNe individually. The median N$_{e}$\sulfurii~ 
of the LISs distribution ($\sim$1700~cm$^{-3}$) is approximately 0.6 lower than that of the PNe rims and shells ($\sim$2700~cm$^{-3}$). An additional way of emphasizing the discrepant electron densities of the two types of nebular components is shown in Fig.~\ref{fig:delta}, top-left panel, which contrasts, per PN, the median N$_{e}$\sulfurii~ of Rims/Shells and LISs. This approach allows to reach exactly the same conclusions that LISs' electron densities are lower than those of rims and shells.

\subsection{Electron temperature} 
Taking into account the electron temperature estimated from the [N{\sc~ii}] diagnostic lines, we observed that there is no significant variation between the two groups (see Fig.~\ref{fig:violinplotTN}). The median values of both distributions differ by $\sim$0.2\%, with overlapping notches. 
The distribution of the \nitrogen~5755/6584 line ratio for the two groups displays a similar range of values from -2 to -1 (in logarithmic scale, see Figure~\ref{UVShmodels}). The two groups also have the same median values (not shown here; -1.70 and -1.71 for the Rims/Shells and LISs, respectively, and comparable lower/upper notches of -1.741/-1.676 for LISs and -1.757/-1.658 for Rims/Shells). Since in the higher ionization structures the N$^{++}$ recombination line could contribute to the [N~{\sc~ii}]~5755~\AA~ auroral line emission, the identity between the two groups suggests that the N$^{++}$ recombination is negligible.

Regarding T$_{e}$[O~{\sc~iii}], we note that the median value of the Rims/Shells group (T$_{e}\sim$10200K) is $\sim$0.88 times lower than that of the LISs (T$_{e}\sim$11600K), whereas the notches of both distributions do not overlap. Therefore, it can be argued that the two groups are statistically different, as far as T$_{e}$[O{\sc~iii}] is concerned.  
A possible interpretation of the different distributions between LISs and Rims/Shells subsets [O{\sc~iii}] temperature could be that the higher electron temperatures in some particular LISs are associated with external heating mechanisms, such as shocks or photoelectric heating by dust grains. For gases with T$_{e}$>8000~K, heating by dust grains is insignificant  due to the electron-grains collisions \citep{Draine1978}, except if large molecules like PAHs are also present \citep{Lepp1988}. The molecular hydrogen emission found in LISs \citep{Akras2017,Akras2020b} points out the presence of an amount of dust which prevents the dissociation of H$_2$ molecules, while PAHs may also be present and responsible for the formation of H$_2$ \citep[e.g.][]{Boschman2015}.

\begin{figure*}
\centering
\includegraphics[width=\columnwidth]{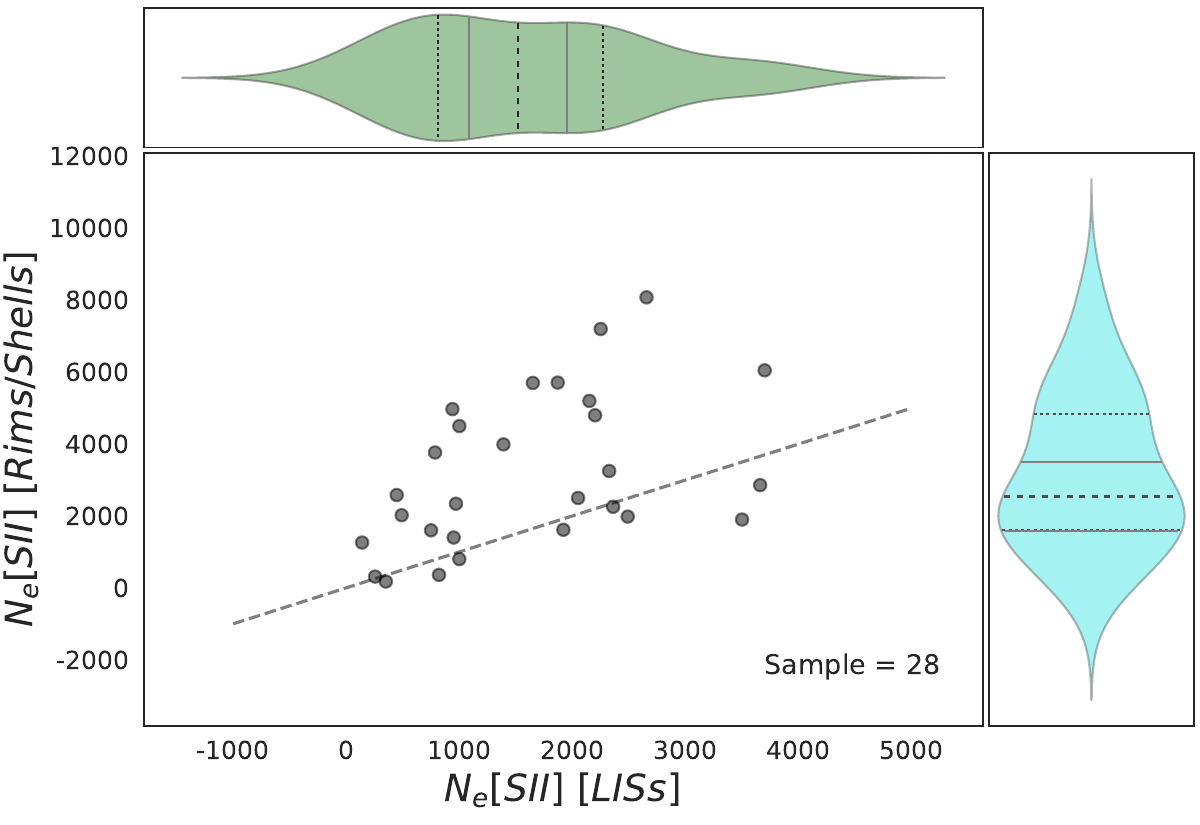}
\includegraphics[width=\columnwidth]{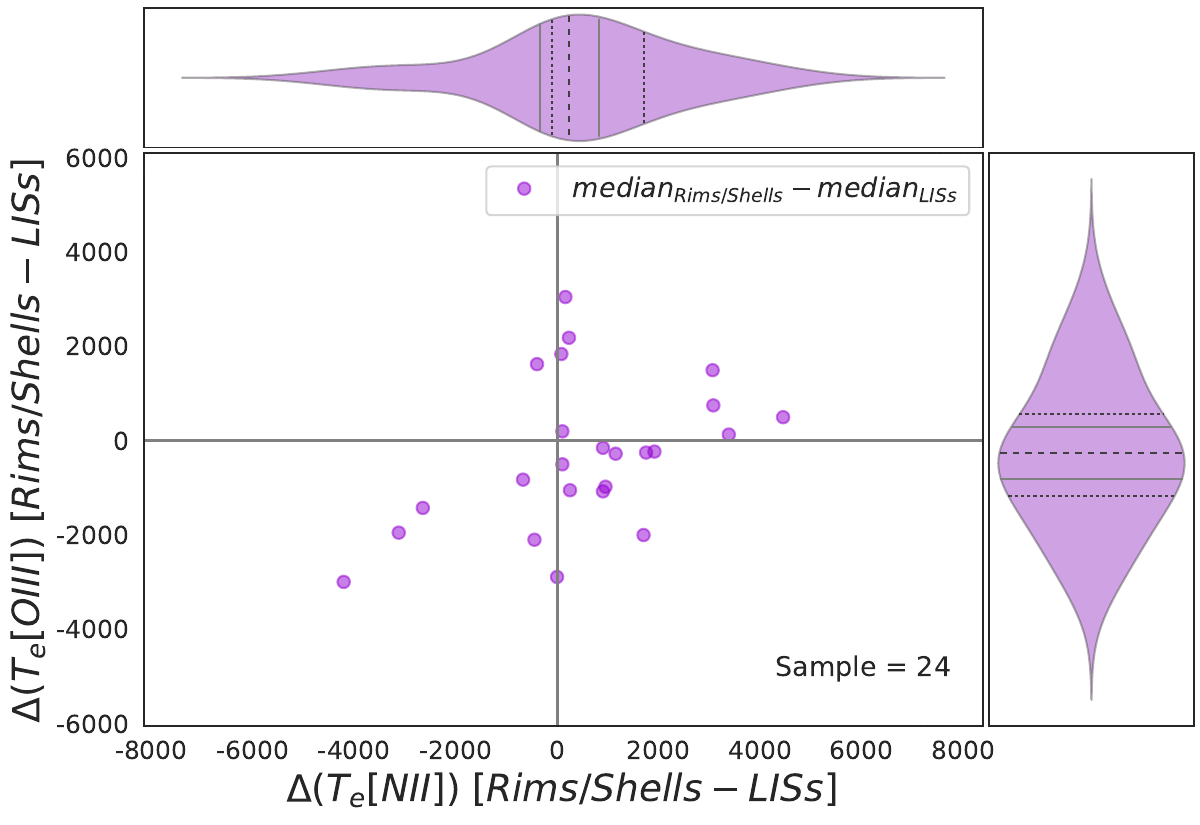}\\
\includegraphics[width=\columnwidth]{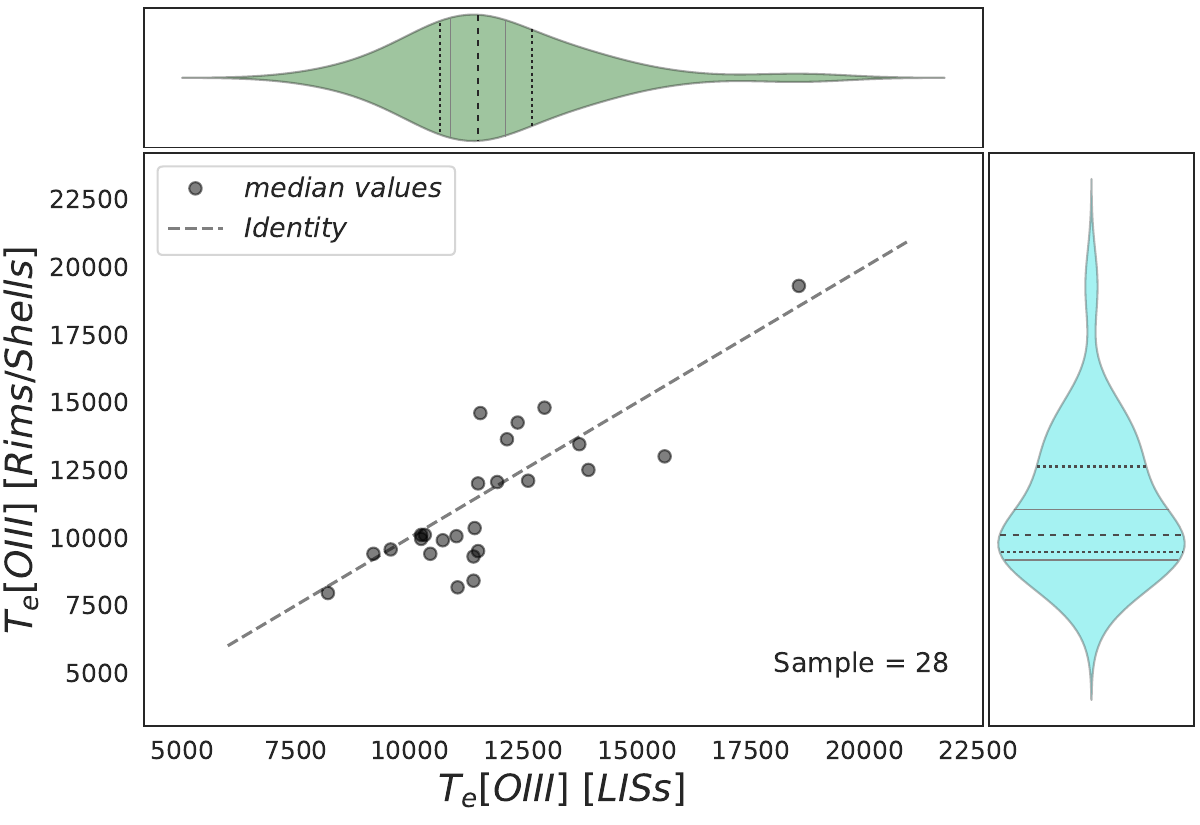}
\includegraphics[width=\columnwidth]{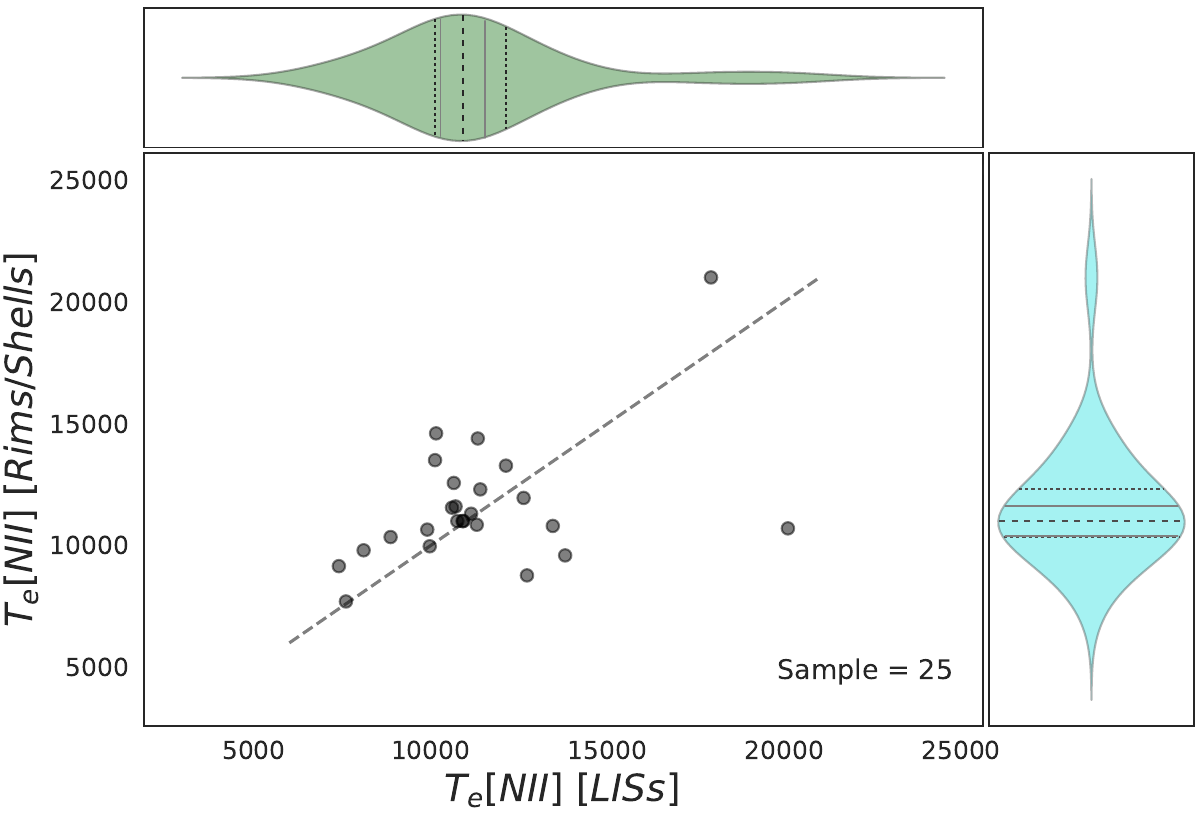}
\caption[]{Comparison of median T$_e$ and N$_e$ 
between Rims/Shells and LISs, for each nebula in the sample, as in Table~\ref{tab:medians}. The dashed-lines represent the identity function.}
\label{fig:delta}
\end{figure*}

The veracity of the above trend for temperatures, can be more stringently tested by the analysis, per nebula, of the median T$_e$ ([O~{\sc iii}] and [N~{\sc ii}]) of the two types of components, as shown in Table~\ref{tab:medians} and Fig.~\ref{fig:delta}. From the top-right panel it can be seen that both differences of electron temperatures estimated through the [N~{\sc ii}] and [O~{\sc iii}] diagnostics exhibit a $\pm$2000~K dispersion centred around zero. This reflects the uncertainties of the estimations, therefore the average $\sim$1400~K higher T$_e$[O~{\sc iii}] of LISs quoted above is within the scatter and cannot be taken as significant. The median values of the two electron temperatures, as in Table~\ref{tab:medians}, are also plotted for Rims/Shells versus LISs in the bottom panels of Fig.~\ref{fig:delta}, which again clearly show that taking the dispersion into account, neither T$_e$[O~{\sc iii}] nor T$_e$[N~{\sc ii}] show significant discrepancies between LISs and Rims/Shells.

\begin{table*}
\caption{Median T$_e$ and N$_e$ for Rims/Shells and LISs, per nebula.} 
\label{tab:medians}
\begin{tabular}{l|ccccccl}
\hline
            & \multicolumn{2}{c}{N$_{e}$[S~{\sc~ii}]} & \multicolumn{2}{|c}{T$_{e}$[N~{\sc~ii}]}     & \multicolumn{2}{c}{T$_{e}$[O~{\sc~iii}]} & \multirow{2}{*}{References} \\           
            
Name        & m$_{\mathrm{Rims/Shells}}$ & m$_{\mathrm{LISs}}$  & m$_{\mathrm{Rims/Shells}}$ & m$_{\mathrm{LISs}}$  & m$_{\mathrm{Rims/Shells}}$ & m$_{\mathrm{LISs}}$ & \\
\hline
NGC~6543    &     4800          &   2200      &     9150          &  7400       &     7950          &   8200    & \cite{1994ApJ...424..800B} \\
NGC~6826    &     800           &   1000      &     7700          &  7600       &     9400          &   9200    & \cite{1994ApJ...424..800B} \\
NGC~7009    &     4500          &   1000      &     9800          &  8100       &     9500          &   11500   & \cite{1994ApJ...424..800B} \\
Hb~4        &     3770          &    790      &     10650         &  9900       &     8550          &   -       & \cite{1997ApJ...487..304H} \\
IC~4634     &     8080          &   2660      &     11000         &  10750      &     9400          &  10450    & \cite{1997ApJ...487..304H} \\
NGC~6369    &     2020          &    490      &     10850         &  11300      &     9300          &  11400    & \cite{1997ApJ...487..304H} \\
NGC~7354    &     2340          &    970      &     11100         &   -         &     9950          &  10250    & \cite{1997ApJ...487..304H} \\
M~2-48      &     1260          &    140      &     10700         &  20100      &     10850         &    -      & \citet{2002AA...388..652L} \\
NGC~7009    &     5700          &   1650      &     11600         &  10700      &     10100         &   10250   & \cite{Goncalves2003} \\
K~4-47      &     1900          &   3500      &     21000         &  17930      &     19300         &   18550   & \cite{Goncalves2004} \\
NGC~7662    &     2500          &   2050      &     14600         &  10150      &     12000         &   11500   & \citet{2004A.A...422..963P} \\
IC~4634     &     5200          &   2150      &     11550         &  10600      &     10050         &   11030   & \citet{2008ApJ...683..272G} \\
He~1-1      &     1600          &    750      &     10800         &  13450      &     12500         &  13930    & \cite{2009MNRAS.398.2166G} \\
IC~2149     &     6050          &   3700      &     12300         &  11400      &     10350         &  11430    & \cite{2009MNRAS.398.2166G} \\
KjPn~8      &      -            &   600	      &       -           &   9030      &       -           &   10330   & \cite{2009MNRAS.398.2166G} \\
NGC~7662    &     3250          &   2330      &     13280         &  12130      &     13450         &  13730    & \cite{2009MNRAS.398.2166G} \\
NGC~7354    &     1980          &   2490      &     14390         &  11330      &     13630         &   12140   & \citet{2010AJ....139.1426C} \\
Necklace    &     360           &    820      &     11000         &  10920      &     14800         &   12960   & \citet{2011MNRAS.410.1349C} \\
ETHOS1      &      -            &     -       &       -           &    -        &       -           &     -     & \citet{2011MNRAS.413.1264M} \\
NGC~3242    &     2860          &   3660      &     13500         &  10120      &     12050         &   11920   & \citet{2013A.A...560A.102M} \\
Hu~1-2      &      -            &     -       &       -           &    -        &       -           &     -     & \citet{2015MNRAS.452.2445F} \\
IC~4846     &     7200          &   2250      &     11950         &  12630      &     9900          &  10730    & \citet{Akras2016}  \\
Wray~17-1   &      180          &    350      &       -           &  12250      &     14250         &  12370    & \citet{Akras2016}  \\
K~1-2       &       -           &    610      &       -           &  9120       &       -           &   14250   & \citet{Akras2016}  \\
NGC~6891    &     1400          &    950      &     9628          &   -         &     9560          &  9580     & \citet{Akras2016}  \\
NGC~6572    &     20840         &   10120     &     12570         &  10650      &     10100         &  10330    & \citet{Akras2016}  \\
M~2-42      &     2580          &    450      &     10340         &  8860       &      -            &   -       & \cite{2016AJ....151...38D} \\
NGC~5307    &      -            &   2950      &      -            &  13090      &      -            &  12330    & \cite{Ali2017}  \\
IC~2553     &      -            &   2400      &      -            &  10790      &      -            &  10930    & \cite{Ali2017}  \\
PB~6        &      -            &   1510      &      -            &  11150      &      -            &  13750    & \cite{Ali2017}  \\
NGC~3132    &      -            &    100      &      -            &  13600      &      -            &   -       & \citet{2020A.A...634A..47M} \\
IRAS        &     1620          &   1920      &     8770          &  12720      &      -            &   -       & \cite{2021arXiv210505186M} \\
Hen~2-186   &     3990 	        &   1390      &     11300         &  11140      &     14600         &  11550    & \citet{2023MNRAS.518.3908M} \\
Hen~2-429   &     5710          &   3760      &     9390          &   -         &     9790          &   -       & \citet{2023MNRAS.518.3908M} \\
IC~4593     &     2260          &   2360      &     9590          &  13800      &     8410          &  11400    & \citet{2023MNRAS.518.3908M} \\
NGC~3918    &     5710          &   1870      &     11000         &  10900      &     12100         &  12600    & \citet{2023MNRAS.518.3908M} \\
NGC~6543    &     4970          &    940      &     9970          &  9970       &     8160          &  11050    & \citet{2023MNRAS.518.3908M} \\
NGC~6905    &      310          &    260      &       -           &    -        &     13000         &  15600    & \citet{2023MNRAS.518.3908M} \\
\hline

\end{tabular} 
\end{table*}

\subsection{Abundances} 
We did not find any statistically significant difference in the chemical abundances of He, O, N, Ne, Ar, S and Cl between the LISs and the high-ionization rims and shells (see Fig.~\ref{fig:correl1} and Table~\ref{tab:violinplot_abun}). 

The linear abundances' ratio correlations 
analyzed also do not show significant differences (slopes and intercepts) 
independently of the LISs' inclusion 
in the correlations (see Table~\ref{tab:correl}). 
Slope and intercepts are similar, taking into account their uncertainties, with and without the LISs. 
The goodness-of-fit, on the other hand, slightly decreases when the LISs are considered. LISs add scatter in the correlations, because of their usually lower S/N ratio in comparison with rims and shells. The correlation found between log(N/H) versus log(N/O) is found to agree, within error, with previous studies \citep[e.g.][]{2013A&A...558A.122G,Akras2016}.
\begin{figure*}
\centering
\includegraphics[width=\columnwidth]{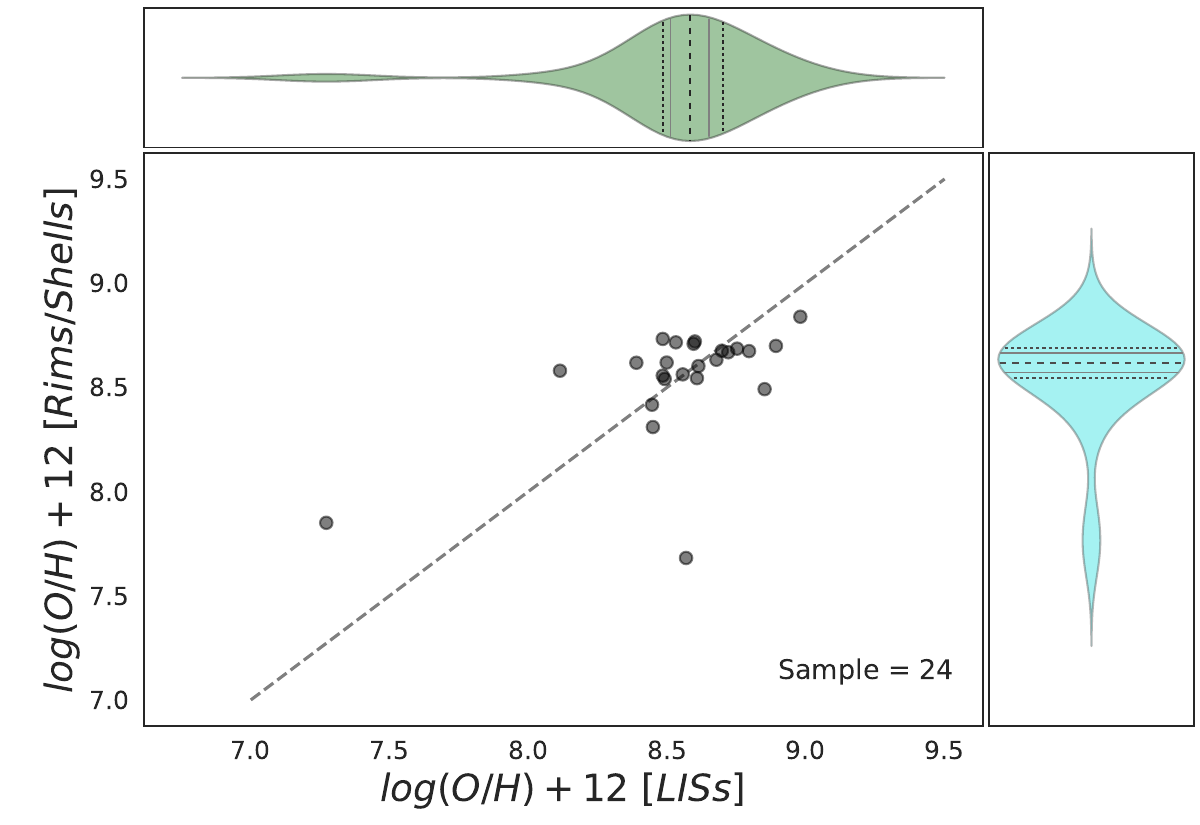}
\includegraphics[width=\columnwidth]{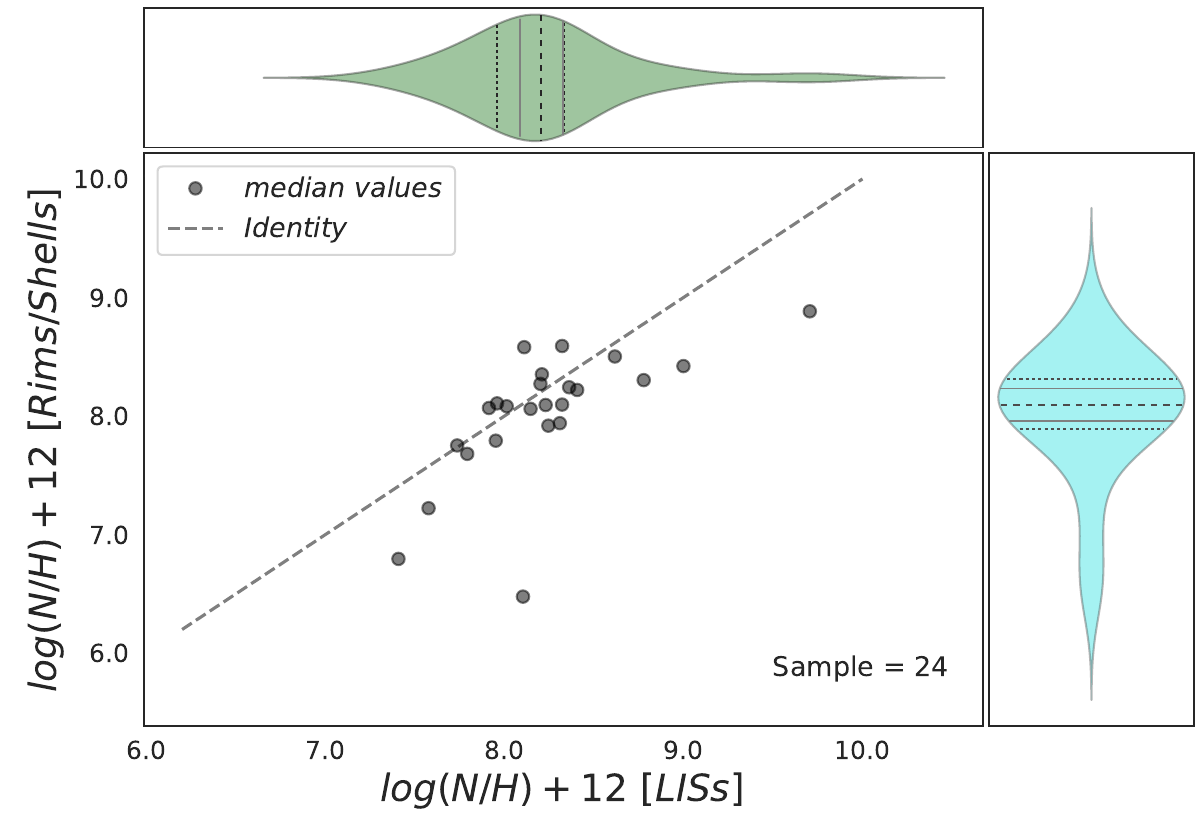}\\
\includegraphics[width=\columnwidth]{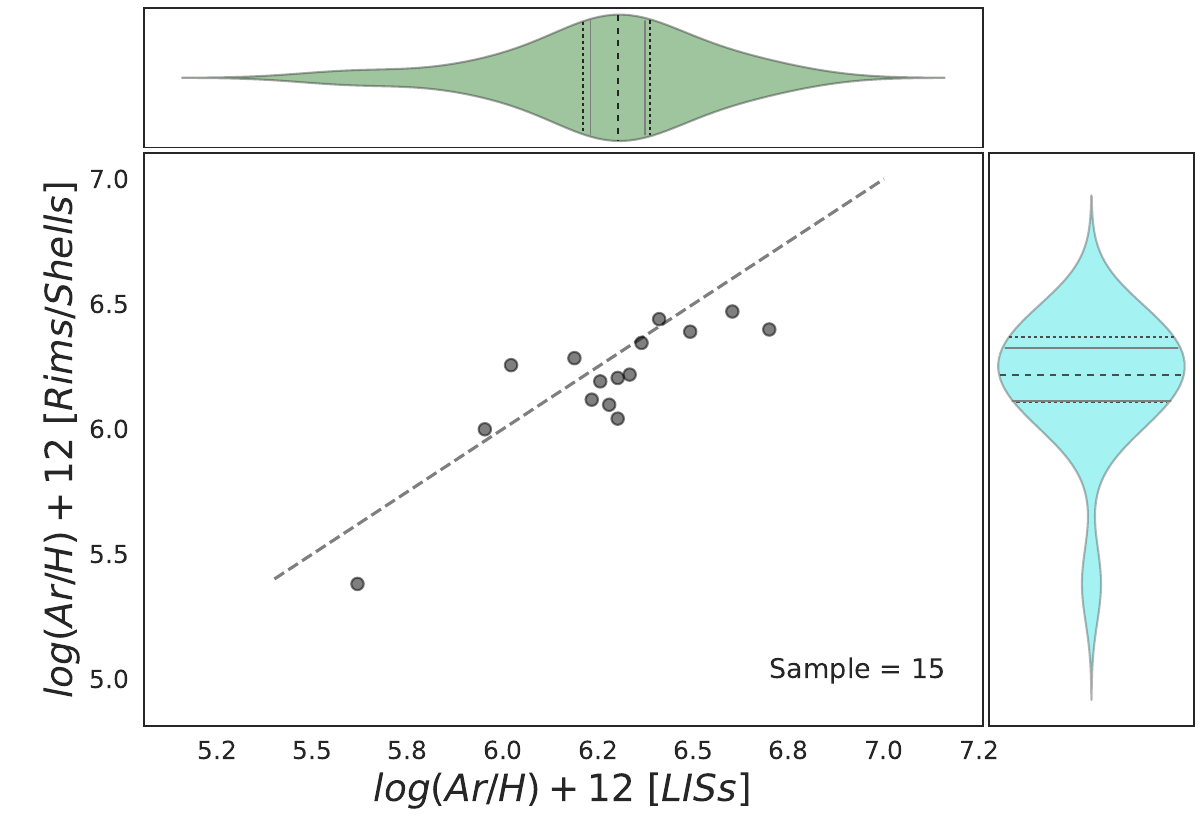}
\includegraphics[width=\columnwidth]{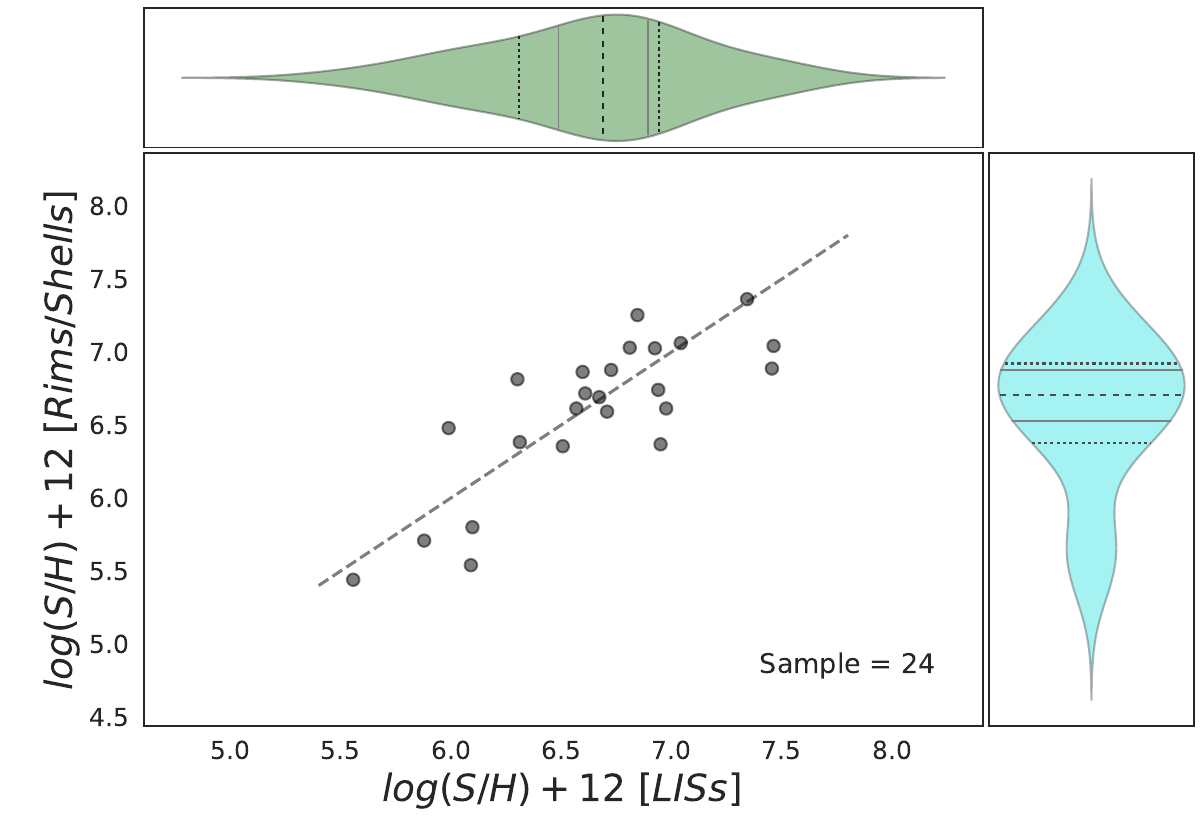}\\
\includegraphics[width=\columnwidth]{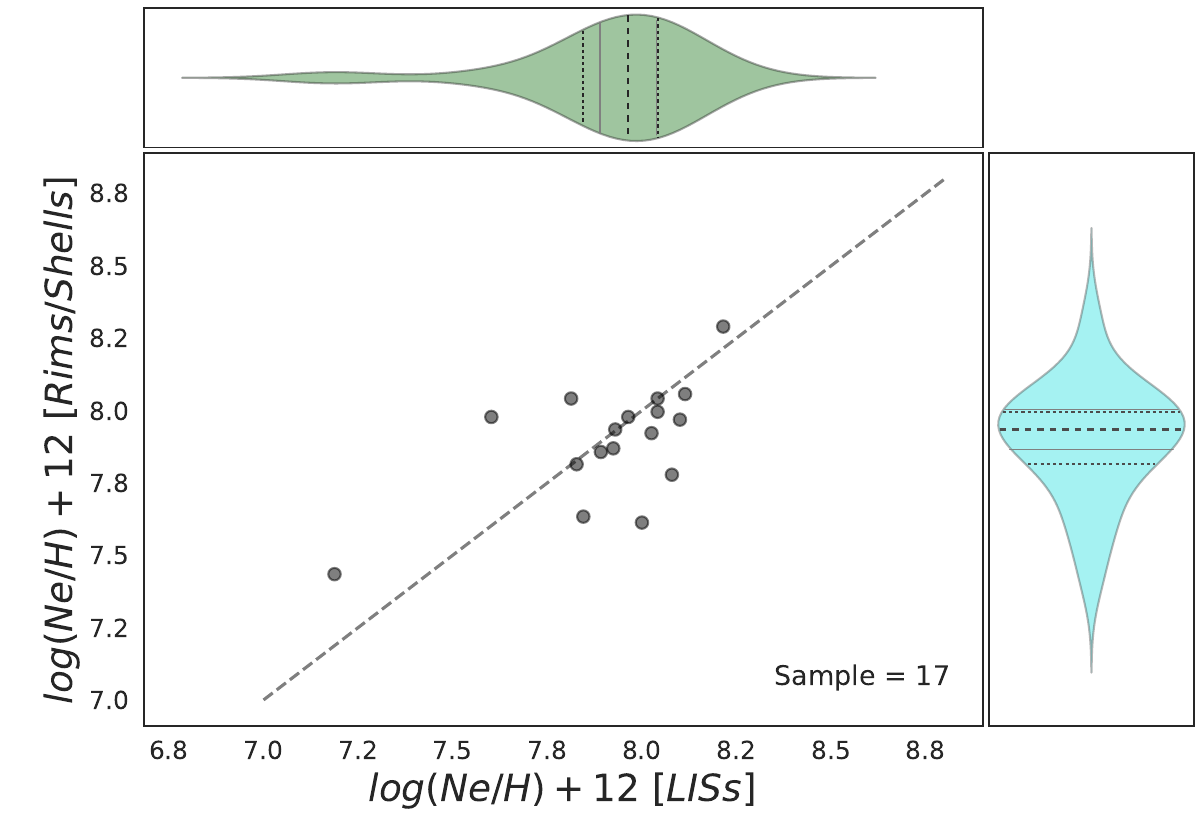}
\includegraphics[width=\columnwidth]{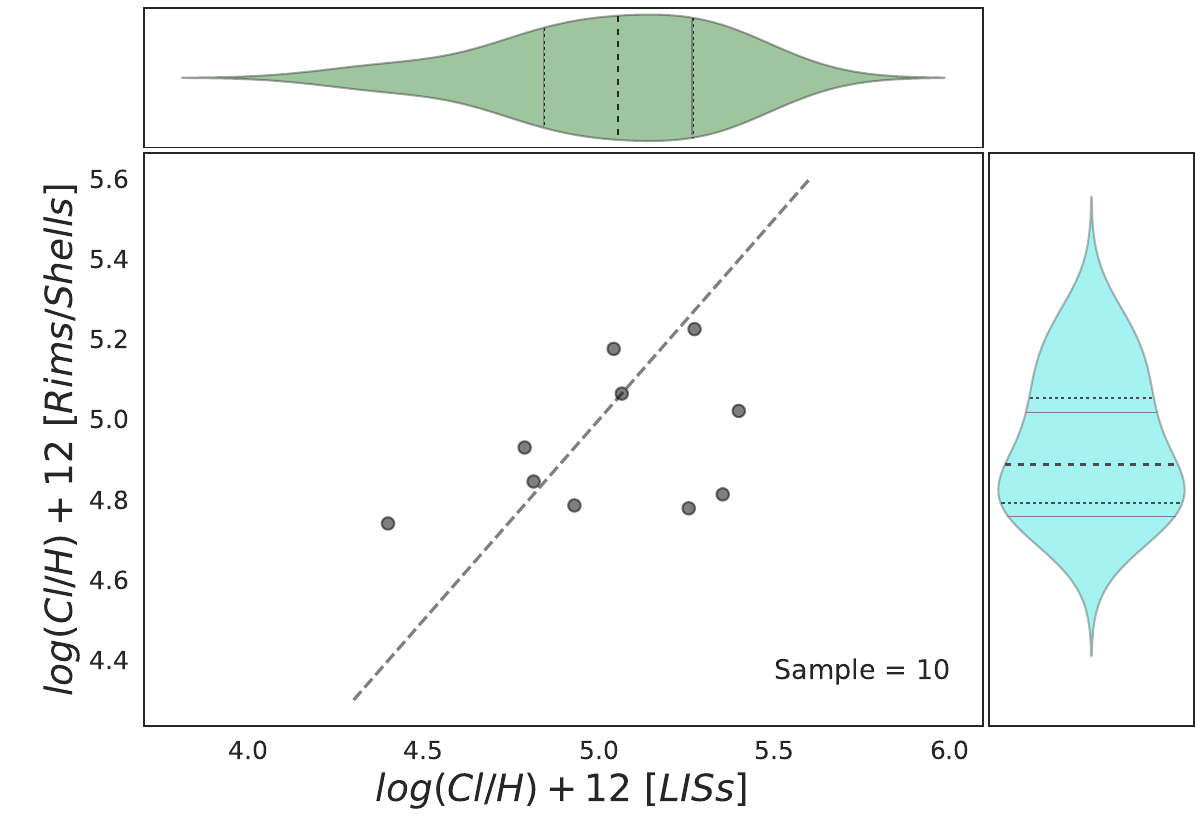}
\caption[]{Same as Fig.~\ref{fig:delta} for total abundances.}
\label{fig:deltaX}
\end{figure*}

As in the cases of T$_e$ and N$_e$, we also examine the veracity of the chemical abundances trends. Figure~\ref{fig:deltaX} shows the distributions of the median values, per chemical element and nebula, for LISs and Rims/Shells. It can be seen that the majority of the 
structures lie near the identity line, although with dispersion. It is important to note that the samples used are small. Nonetheless, it is verified that the abundances between the two groups do not differ significantly, as previously established (see Fig~\ref{fig:correl1}).

\subsection{Excitation mechanism} 
SMB's and BPT's emission-line ratio diagrams were used in an attempt to found any significant difference between the excitation processes dominating Rims/Shells and LISs. From SMB, it is observed that most LISs are in the PNe region, with higher [N~{\sc~ii}]/H$\alpha$ and [S~{\sc~ii}]/H$\alpha$ ratios compared to the Rims/Shells. From their violinplots distributions and corresponding statistical parameters, we verified that there is no overlap between the two groups. Concerning the BPT, we found that LISs and Rims/Shells exhibit different median [O~{\sc~iii}]~$\lambda$5007/H$\beta$~face values, with the latter slightly lower, while their notches do overlap. As for the [O~{\sc~iii}]~$\lambda$4363/H$\beta$ ratio, the statistics shows similar medians and clear overlapping.  Motivated by the large [O~{\sc~iii}]~$\lambda$4363/H$\beta$ ratios found in LISs, we also explore the high-ionization line ratio He{\sc~ii}~$\lambda$4686/H$\beta$. Despite, LISs have lower median values than rims and shells, there is a number of LISs with He~{\sc~ii}~$\lambda$4686/H$\beta$~between 0.7-1 (e.g. Wray17-1, K~1-2, Hu1-2, Necklace among others). This particular subgroup of PNe display complex morphologies with highly-collimated or jets or jet-like structures. Furthermore, the host nebulae of this subgroup of LISs also exhibit high He{\sc~ii}~$\lambda$4686/H$\beta$.

An analysis of the two groups based on the [O~{\sc~ii}]$\lambda$3727/$\lambda$5007 versus [O~{\sc~i}]$\lambda$6300/H$\alpha$ diagram was also performed, and a clear separation between the Rims/Shells and LISs is observed. LISs occupy the top-right corner, with high values of [O~{\sc~ii}] and [O~{\sc~i}], while the Rims/Shells are located in the bottom-left corner. This result is also verified with the split violinplots, where their medians differ at a 95 percent confidence level (see Table~\ref{tab:vio1}). Combining emission lines of the same element (oxygen), from three different 
ionization states, we 
highlight the different ionization state of LISs and host nebulae, 
avoiding the effect of the chemical abundance of the combination of different elements, as in the previous 
diagrams.

\subsection{Model predictions}
The intriguing characteristics of low-ionization structures relative to their host PNe has been  calling astronomers' attention for several years regarding their origin and the dominant excitation mechanism (photoionization and shock-heating processes). Their usually stronger low-ionization lines relative to \ha~(e.g., \nitrogen~6548+6584/\ha, \sulfurt~6716+6731/\ha, \oxygeni~6300/\ha) have been attributed either shocks \citep[e.g.][]{Hartigan1994,Dopita1997,Goncalves2004,Akras2016} or UV photoionization process \citep[e.g.][]{1997ApJ...487..304H,Goncalves2003,Ali2017}.

The recent discoveries of molecular hydrogen (H$_2$) associated with LISs \citep{2015MNRAS.452.2445F,Akras2017,2018ApJ...859...92F,Akras2020b} have entailed the presence of highly dense gas (>10$^{4-5}$~cm$^{-3}$) to shelf-shield the molecular component and prevent its dissociation. Such high-density structures are able to produce strong low-ionization lines similar to photo-dissociation regions (PDRs) or low-ionization nebulae (low logU). 

To further investigate the dominant mechanisms in LISs, we compared the observations with the predictions from photoionization and shock models. The regions occupied by LISs, Rims/Shells and the distribution of fast/slow shock models and photoionization models are presented in eight emission line DDs in Fig.~\ref{UVShmodels}. At first look, we find a very good match between the observations and the regime of photoionization models (blue contour) in all DDs, but shock can not be easily ruled out as several line ratios can also be reproduced by fast- or slow-shock models.

It should be noted that Rims/Shells and LISs show a different slope in the \oxygenii/\oxygeniii~versus~\oxygeni/\ha~DD (see Fig~\ref{fig:dds}), indicating a different ionization state in LISs relative to their host PNe, such as a mini-PDR around a dense molecular core.

Concerning the widely used SMB -- \ha/\nitrogen~6548+6584 versus \ha/\sulfurt~6716+6731 -- diagnostic diagram \citep{Sabbadin1977}, we demonstrate that LISs lie in the bottom-left corner of the PNe locus, close to the locus of observed SNRs. Besides low-ionization models (logU$<$-3), only shock with velocities $<$100~km~s$^{-3}$ are able to produce \ha/\nitrogen~and \ha/\sulfurt~6716+6731 line ratios similar to LISs, but fail to reproduce other lines.

LISs are found to exhibit a systematic higher T$_{e}$\oxygeniii~than Rims/Shells but comparable T$_{e}$\nitrogen~(see Fig.~\ref{fig:violinplotTN} and Table~\ref{tab:violinplot}). A statistical analysis on the temperature sensitive diagnostic line ratios has also been performed. LISs are characterized by a median  log(\oxygeniii~$\lambda$4363/$\lambda$5007)=-1.9505 higher than the median of rims and shells (-2.0303). This deviation is statistically significant if we take into account the lower/upper notches (LISs: -1.9989/-1.9021, Rims/Shells:-2.1236/-1.937), which demonstrate the 95 percent of the confidence interval (CI) for the median values. 
 
The distribution of the LISs and Rims/Shells subsets in the \oxygeniii~$\lambda$4363/$\lambda$5007 versus \nitrogen~$\lambda$5755/$\lambda$6584 DD is presented in Fig.~\ref{UVShmodels}. Most of the data points lie well within the regime of photoionization models. There is, though, a distinct small group of LISs which display an enhanced \oxygeniii~$\lambda$4363/$\lambda$5007 ratio relative to the predicted ratio from photoionization models, lying in an area where only low-velocity shock models are found. This particular group of LISs includes the following PNe: Hen~2-186 (1 LIS), NGC~3918 (2 LISs), NGC~6543 (1 LIS), K~1-2 (3 LISs), NGC~6572 (1) Hen 1-1 (1 Rim/Shell), KjPn8 (1 LIS), and NGC~7009 (1 LIS). We should also mention that log(\oxygeniii~$\lambda$4363/$\lambda$5007)>-1.5 seems to be a good indicator of shock-heated gas. \cite{Leung2021} also came to a similar conclusion based on AGN and shock models. More specifically, AGN models cannot reach log(\oxygeniii~$\lambda$4363/$\lambda$5007) higher than -1.5 for any log(U) value, while shock models yield values between -2 and -1, in agreement with our results \citep[see figures 12 and 13 in][]{Leung2021}. It is worth mentioning that the criterion log(\oxygeniii~$\lambda$4363/$\lambda$5007)>-1.5 is valid only for environments with density $\leq$7$\times$10$^5$~cm$^{-3}$ (critical density of the \oxygeniii~5007\AA~line). In the case of a denser gas, the \oxygeniii~5007\AA~line will be collisionally de-excited and will result in a high \oxygeniii~$\lambda$4363/$\lambda$5007 ratio resembling shock-heated gas.

\section{Conclusions}\label{sec6}
The main conclusions extracted from the physical, chemical and excitation properties of the largest sample of LISs, rims and shells of PNe analyzed so far are listed below. 
\begin{itemize}
 \item LISs are statistically different from Rims/Shells in terms of N$_{e}$\sulfurt. The former exhibit $\sim$2/3 lower electron density ($\sim$1700~cm$^{-3}$) than the latter ($\sim$2700~cm$^{-3}$) components. 
 \item Though LISs have median T$_{e}$\oxygeniii~ comparable with those of the Rims/Shells, respectively $\sim$11600K and $\sim$10200~K (both with large dispersion), the distribution \oxygeniii~ temperatures also has a well-marked bimodality, not easily explained. The \nitrogen~ electron temperatures show no difference between the two types of nebular components, LISs and Rims/Shells, with a median value of $\sim$10800~K. 
 \item No statistical difference in the chemical composition is found between LISs and Rims/Shells, based on the analysis of helium, nitrogen, oxygen, neon, argon, chlorine and sulphur. 
 \item Shock models with low-velocity shocks, as well as photoionization modelling of PNe with low ionization parameter are both able to produce the line ratios found in LISs.
 \item The [N~{\sc~ii}]/[S~{\sc~ii}]$\leq$0.25 ratio is found to distinguish fast-shock ($>$100~km~s$^{-1}$) models from photoionization and slow-shock models. 
 \item The diagnostic diagram of the temperature dependent ratios shows that  log(\oxygeniii~4363/5007)$>$-1.5 is a good tracer of shock-heated gas, for electron densities $<$7$\times$10$^5$~cm$^{-3}$. 
 \item The vast majority of LISs and Rims/Shells have line ratios in agreement with the predictions of the photoionization models, yet there are a few LISs for which shocks could be present.
 \item Individual line ratios are not adequate to distinguish photo-heated and shock-heated gas, a combination of them provides a more robust separation.
\end{itemize}

\section*{Acknowledgements}
We would like to thank the anonymous referee for her/his careful reading of our work and helpful comments/suggestions, which have helped to improve the paper.
This research is support by a PhD grant from CAPES -- the Brazilian Federal Agency for Support and Evaluation of Graduate Education within the Education Ministry. SA acknowledges support under the grant 5077 financed by IAASARS/NOA. DGR acknowledges the grants 313016/2020-8 (CNPq) and 200.527/2023 (FAPERJ).

\section*{Data Availability}

The two grids of photoionization and shock models underlying this article are available in the \href{http://3mdb.astro.unam.mx/}{ Mexican
Million Models database}.





\typeout{}
\bibliographystyle{mnras}
\bibliography{example} 






\appendix

\section{Data visualization}\label{apA}

A representative -- and classical -- way to explore large datasets is through the use of histograms in order to determine the distribution of each feature under investigation. Alternatively, the use of diagrams such as \textit{boxplots} \citep[]{1977eda..book.....T} provides a better visualization for the quantities such as the quartiles, the median, the interquartile range (IQR) and also the outliers. The definition of these quantities is given below.

\begin{itemize}
    \item \textit{Quartiles} specify the location of the 25$^{th}$ (Q1) and 75$^{th}$ (Q3) percentiles. The 25$^{th}$ percentile corresponds to the 25 percent of the values that are less than or equal to this value. A similar definition for the 75$^{th}$ percentile;
    \item \textit{Median} represents not only the mid-point of the distribution but also the 50$^{th}$ (Q2) percentile;
    \item \textit{Interquartile range (IQR)} is the distance/range between the third and first quartiles;
    \item The minimum \textit{(min)} and maximum \textit{(max)} value of the sample, excluding the outliers, which are determined for the distance between the first and third quartiles and 1.5 times the IQR\footnote{This is related to a characteristic of the Normal Distribution, the 1.5 in minimum and maximum value ends up being approximately $\pm$2.7$\sigma$ (being $\sigma$ the standard deviation) from the mean, which corresponds to a 99.3 percent of the data for a normal distribution. Then, any data point lower than (or greater than) the min (max) is considered as an outlier.};
    \item \textit{Notches} 
    Another very important parameter for a statistical approach is the \textit{notches} \citep{2017..book.....C}. The \textit{notches} are related with the median, IQR and the number of observations/population (\textit{n}) of each subset, and they are used to demonstrate the 95 percent of confidence interval (CI) for the median value: $m\pm 1.58\times IQR / \sqrt{n}$. When the notches of two samples do not show an overlap, the medians of each of the distributions are considered significantly different \citep{2014naturemethods}. A possible overlap does not necessarily rule out the possibility that the two samples are different.
\end{itemize}

\begin{figure}
    \centering
    \includegraphics[width=0.96\columnwidth]{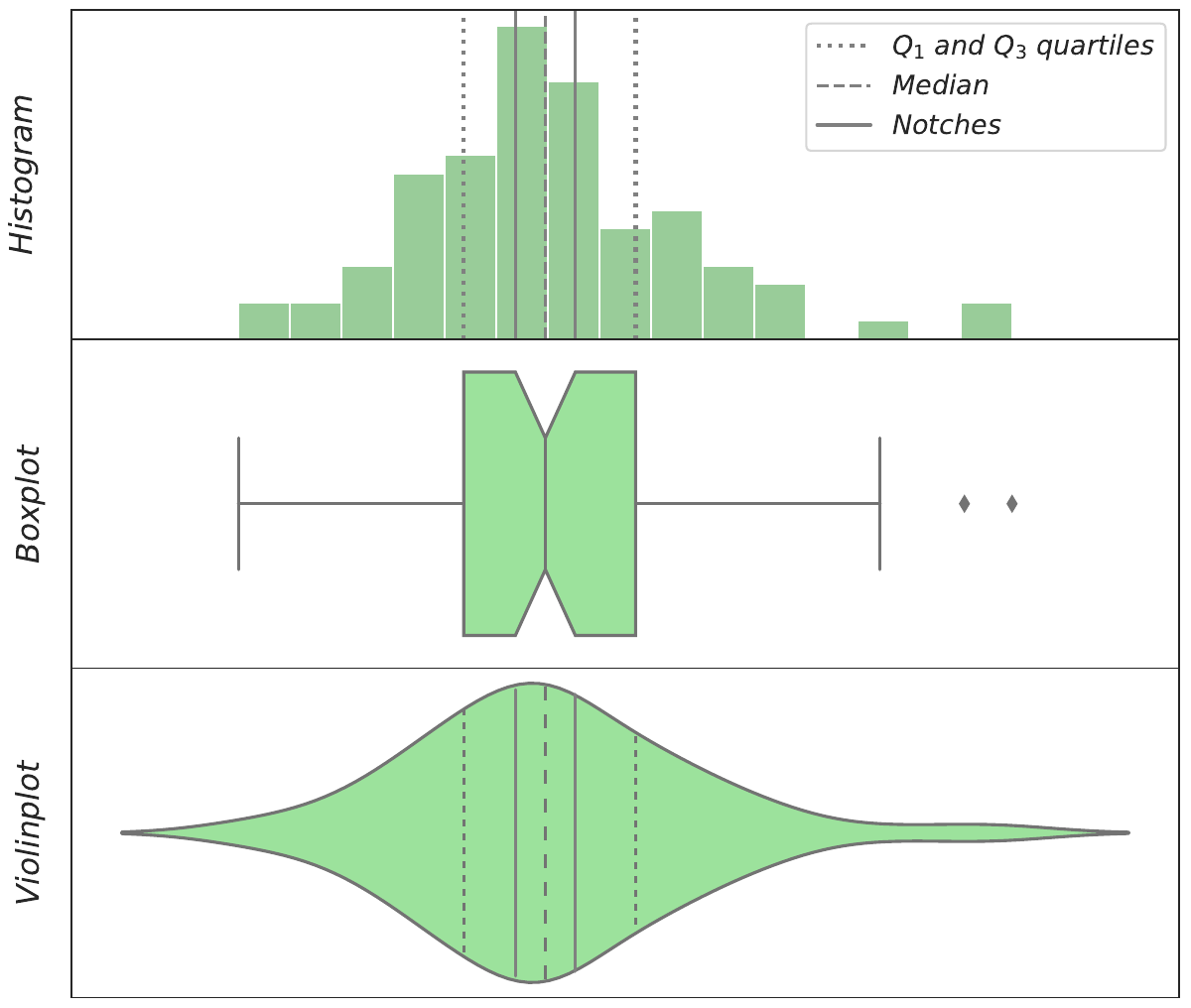}
    \caption{Distribution of a data sample for three different visualization plots: histogram, boxplot and violinplot. Q$_{1}$ and Q$_{3}$ quartiles represent the location of the 25$^{th}$ and 75$^{th}$ percentiles, respectively, and they are illustrated by vertical dash-dotted lines. The median values represent the 50$^{th}$ percentile, which is also the mid-point of the distribution, and it is shown by a vertical dashed line. The notches show the most likely values expected for the median, and they are represented by vertical solid lines. In the case of the boxplot, the actual structure of the boxes coincide with the aforementioned parameters. The violinplot also demonstrates the density distribution of a sample, where wider (narrower) regions represent a higher (lower) probability that members of the population adopt the given value.}
    \label{fig:violinplot}
\end{figure}

A second --and more integrated-- visualization approach is through the use of \textit{violinplots}, which combine a boxplot and kernel density estimation together in one diagram \citep[]{10.2307/2685478}. This representation of a data sample provides information about the shape of the distribution, such as their peaks and their positions, or even unveil the presence of clustering in the data (e.g. a bimodal distribution). Moreover, having the density distribution in the violinplots, it can be seen that the wider the section (the size of the violin in the y-axis of Figure~\ref{fig:violinplot}, bottom panel) the higher that probability to get the corresponding value, whereas a narrower section represent a lower probability. Figure~\ref{fig:violinplot} illustrates these parameters on a violinplot, together with a boxplot and histogram, for comparison purposes. 

\section{3MdB database}\label{apB}

The grid of photoionization models in the 3MdB was constructed using {\sc~cloudy} v17.01 \citep{Ferland2017} and covers a wide range of physical parameters: T$_{\rm eff}$ and L (or equivalently log(U)), chemical abundances, density and size \citep[see ][]{2014MNRAS.440..536D}. The entire grid consists of 724,386 models, but only 116,121 of them are used that satisfy the criteria (flag com6=1) in \cite{2014MNRAS.440..536D}.

The stellar parameters of effective temperature (T$_{\rm eff}$) and luminosity (L) of PNe central stars are crucial as they define the energy distribution of the ionizing photons responsible for the ionization and excitation of atomic gas. Hence, it is worth to explore the emission line ratios as functions of stellar T$_{\rm eff}$ and L. The grid of photoionization models in the 3MdB covers a range T$_{\rm eff}$ from 25 up to 300~kK for a black-body approximation, and from 50 to 180~kK using the atmosphere stellar models from Rauch \citep{Rauch2003}, 2$\times$10$^2<$L$<$1.78$\times$10$^4$~L$\odot$ and a wide range of density from 30 up to 3$\times$10$^5$~cm$^{-3}$. No noticeable differences in the line ratios between the blackbody and atmosphere stellar models is found. 

Photoionization models' entanglement with T$_{\rm eff}$ and L compels us to explore the more general value of the ionization parameter (logU) which encompass both stellar parameters, and it is more suitable and widely used. It is defined as the ratio of ionizing photon density to gas density multiplied by speed of light at a distance r from the central source (U(r)=Q/4$\pi$r$^2$n$_{\rm H}$c), where Q is the rate of ionizing photons and directly related with the stellar parameters, n$_{\rm H}$ the hydrogen density and c is the speed of light. This dimensionless parameter can easily distinguish regions that are UV-dominated or not, and it varies from -1 to -5.

Finally, four metallicities/abundances sets are available in the 3MdB: log(O/H)=-3.66, -3.36, -3.06 and -2.76 \citep[see also ][]{2014MNRAS.440..536D}.

\begin{figure*}
\centering
\includegraphics[width=18cm]{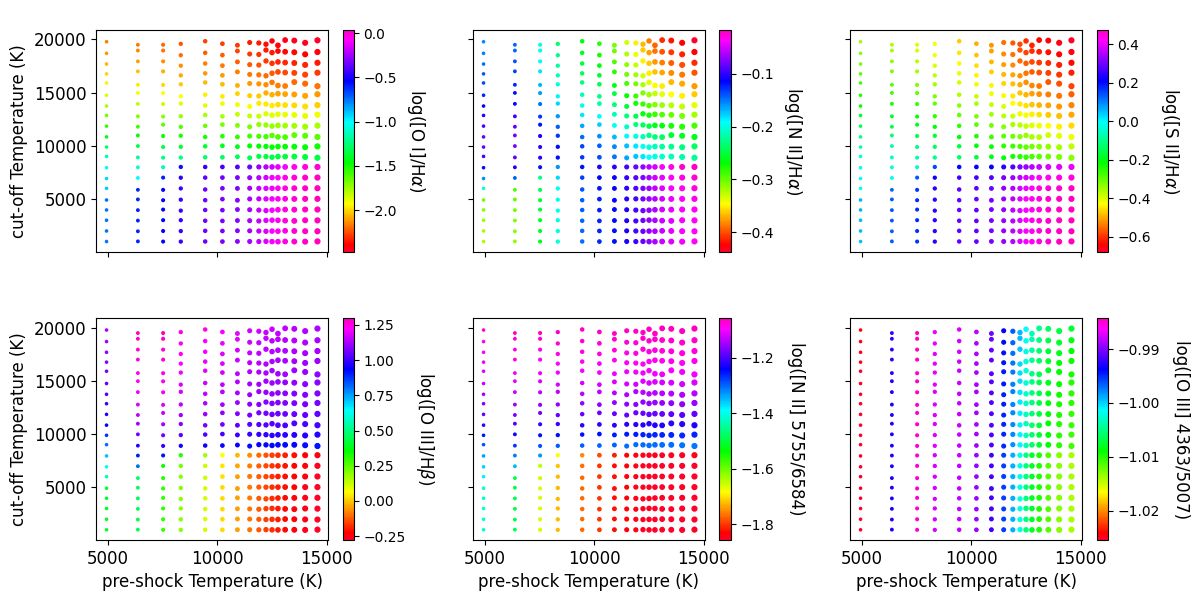}
\caption[]{T$_{\rm pre-shock}$ versus T$_{\rm cut-off}$ plots of emission line ratios for the high velocity incomplete models (namely Allen2008\_cut). The colour of the symbols correspond to the values of each line ratio and their size to the shock velocity. The larger the points, the higher the shock velocity. Only the models with shock velocity $<$500~km~s$^{-1}$ and pre-shock transverse magnetic fields $<$0.5$\mu$G are plotted.}
\label{Tpsh_Tco_AD}
\end{figure*}

The shock models were generated using the code {\sc mappings v}, version 5.1.13 \citep{Sutherland2017,Sutherland2018}. The total number of available shock models in the 3MdB is 199,750 and a wide range of physical parameters is covered such as the shock velocity, the pre-shock and cut-off temperatures, the ionization state of the pre-shocked gas, the pre-shock density and metallicity, and the pre-shock transverse magnetic field. For more details, we refer the reader to \citep[][]{Dopita1995,Dopita1997,Allen2008}.

Five different metallicities/abundances sets (LMC, SMC, solar, twice-solar and one namely Allen2008\_Dopita2005) were used for the total grid of high velocity shock models and only one abundance set (namely 3MdB-PNe2014-solar) for the low velocity shock models.

The pre-shock (T$_{\rm pre-shock}$) and cut-off temperatures (T$_{\rm cut-off}$) are among the most important parameters in shock models. At this point, it is noteworthy to define the complete and incomplete or truncated shock models. Complete shock models are like those from \citet{Allen2008} and they are defined as the shock models for which T$_{\rm cut-off}=$1000~K. Any change in T$_{\rm cut-off}$ results in different spatial extent of the post-shock regions and consequently to the integrated emission and line ratios.

T$_{\rm cut-off}$ is the equivalent of T$_e$ stopping criterion (4000~K) in the {\sc~cloudy} photoionization models \citep{Ferland2017, Bohigas2008}. Both define the temperature of the gas at a certain distance behind the shock (or from the central star) at which the model must stop, as not significant emission is considered to emanate from these regions since the gas has cooled down and totally recombined. For the incomplete/truncated shock models in the 3MdB, T$_{\rm cut-off}$ varies from 1000 up to 20000~K, and it has a significant impact on the resultant emission line spectra \citep[e.g.][]{Alarie2019,AlarieDrissen2019}.

The temperature of the pre-shocked gas (T$_{\rm pre-shock}$) has also an important effect on the predictions of shock models. In particular, shocks with velocities higher than 100 km~s$^{-1}$ can significantly alter the physical conditions and ionization structure of the pre-shocked gas, resulting in a wide range of shock spectra. On the other hand, shocks with velocities $<$75~km~s$^{-1}$ do not have any significant influence on the pre-shocked gas. Therefore, the grid of low shock velocity models has been built considering a priori different ionization fractions for the pre-shocked gas. 

In Figure~\ref{Tpsh_Tco_AD}, we displays the dependence of various emission line ratios as functions of T$_{\rm pre-shock}$, T$_{\rm cut-off}$ and shock velocity for the high velocity incomplete grid of models (namely Allen2008\_cut). Shock velocity is constrained to $<$500~km~s$^{-1}$ (and T$_{\rm pre-shock}<$15000~K) and the pre-shock transverse magnetic field <0.5~$\mu$G. The size of the symbols represent the velocity of the shock.

The [O~{\sc i}]/\ha, [N~{\sc ii}]/\ha~ and [S~{\sc ii}]/\ha~line ratios decrease for higher T$_{\rm cut-off}$, while [O~{\sc iii}]/\hb~increases. Moreover, it should be pointed out that [O~{\sc i}]/\ha, [N~{\sc ii}]/\ha~ and [S~{\sc ii}]/\ha~ increase (or decrease) for higher shock velocity depending on the T$_{\rm cut-off}$ parameter. For T$_{\rm cut-off}>$10000~K, the ratios decrease while for T$_{\rm cut-off}<$10000~K the ratios increase. As for [O~{\sc iii}]/\hb, it also increases as function of shock velocity but only in shock models with T$_{\rm cut-off}<$10000~K. For higher T$_{\rm cut-off}<$, the [O~{\sc iii}]/\hb~ratio seems to be independent.

The \nitrogen~5755/6584 and \oxygeniii~4363/5007 ratios are also presented in Figure~\ref{Tpsh_Tco_AD}.
The former shows no correlation with T$_{\rm pre-shock}$ or shock velocity but it become higher for increasing T$_{\rm cut-off}$. The latter takes values in a very narrow range from -0.98 to -1.03 being nearly unaffected by the T$_{\rm pre-shock}$ and T$_{\rm cut-off}$ parameters. 

Pre-shock density is also a crucial parameter in shock models. The grid of low velocity models \citep{Alarie2019} was built considering four pre-shock densities: 10, 100, 1000 and 10000~cm$^{-3}$, whereas the grid of high velocity shock models \citep[complete and incomplete; ][]{Allen2008} is constructed for only one pre-shock density (1~cm$^{-3}$). Only, the pre-shock density in the shock models with solar metallicity ranges from 0.01 up to 1000~cm$^{-3}$. 

Last but not least, the pre-shock transverse magnetic field also has an important impact on the spectra and emission line ratios. It is found that the affect of pre-shock transverse magnetic field on the emission lines is more complicated. For the models with T$_{\rm pre-shock}\leq$13000~K and any value for T$_{\rm cut-off}$, all the aforementioned ratios -- [O~{\sc i}]/\ha, [N~{\sc ii}]/\ha, [S~{\sc ii}]/\ha, [O~{\sc iii}]/\hb, \nitrogen~5755/6584 and \oxygeniii~4363/5007 -- appear unaffected by the magnetic field. For the models with T$_{\rm pre-shock}\geq$13000~K and T$_{\rm cut-off}\geq$13000~K, the [O~{\sc i}]/\ha, [N~{\sc ii}]/\ha~ and [S~{\sc ii}]/\ha~line ratios decrease for higher magnetic field while the [O~{\sc iii}]/\hb, \nitrogen~5755/6584 and \oxygeniii~4363/5007 ratios are almost invariable. For the cases of T$_{\rm pre-shock}\geq$13000~K and T$_{\rm cut-off}\leq$13000~K, [O~{\sc i}]/\ha, [O~{\sc iii}]/\hb~and \nitrogen~5755/6584 do not show significant changes, [N~{\sc ii}]/\ha~ increases and [S~{\sc ii}]/\ha~ and \oxygeniii~4363/5007 decrease for higher magnetic field. All these changes in the line ratios with the magnetic field become readily apparent in the models with B$>$4$\mu$G.

It is worth to clarify that the \oxygeniii~4363/5007 temperature diagnostic becomes as low as -1.2 (in logarithmic scale) only for high velocities models, T$_{\rm pre-shock}\geq$13000~K, T$_{\rm cut-off}\leq$13000~K and B$>$4$\mu$G. For the rest of the models, the \oxygeniii~4363/5007 ratio is nearly to -1.1~.


\bsp	
\label{lastpage}
\end{document}